\documentclass[aps,physrev,preprint,groupedaddress,
nofootinbib]
{revtex4-2}

\usepackage{amsmath}
\usepackage{booktabs}
\usepackage{longtable}
\usepackage{threeparttablex}
\usepackage{graphicx}
\usepackage{multirow}
\usepackage{hyperref}
\usepackage{color}
\usepackage{listings}
\newcommand{\mathbi}[1]{\boldsymbol{#1}}
\begin{document}


\title{\texttt{PHECT}: A lightweight computation tool for pulsar halo emission}


\author{Kun Fang}
\email[]{fangkun@ihep.ac.cn}
\affiliation{Key Laboratory of Particle Astrophysics, Institute of High Energy Physics, Chinese Academy of Sciences, Beijing 100049, China}


\date{\today}

\begin{abstract}
$\gamma$-ray pulsar halos, most likely formed by inverse Compton scattering of electrons and positrons propagating in the pulsar-surrounding interstellar medium with background photons, serve as an ideal probe for Galactic cosmic-ray propagation on small scales (typically tens of parsecs). While the associated electron and positron propagation is often modeled using homogeneous and isotropic diffusion, termed here as standard diffusion, the actual transport process is expected to be more complex. This work introduces the Pulsar Halo Emission Computation Tool (\texttt{PHECT}), a lightweight software designed for modeling pulsar halo emission. \texttt{PHECT} incorporates multiple transport models extending beyond standard diffusion, accounting for different possible origins of pulsar halos. Users can conduct necessary computations simply by configuring a \texttt{YAML} file without manual code edits. Furthermore, the tool adopts finite-volume discretizations that remain stable on non-uniform grids and in the presence of discontinuous diffusion coefficients. \texttt{PHECT} is ready for the increasingly precise observational data and the rapidly growing sample of pulsar halos.
\end{abstract}


\maketitle

\section{Introduction}
\label{sec:intro}
Pulsar halos are extended $\gamma$-ray sources around middle-aged pulsars (see Ref.~\cite{Fang:2022fof,Liu:2022hqf,Lopez-Coto:2022igd,Amato:2024dss} for reviews), initially detected in the TeV regime \cite{Abdo:2007ad,Abeysekara:2017old} and hence also referred to as TeV halos \cite{Linden:2017vvb}. These structures are most likely produced by relativistic electrons and positrons\footnote{Hereafter, \textit{electrons} refers to both electrons and positrons unless explicitly distinguished.} that have escaped from pulsar wind nebulae (PWNe) into the interstellar medium (ISM), where they generate $\gamma$ rays via inverse Compton scattering (ICS) with background photons. The background photon field can be regarded as homogeneous over the typical scale of pulsar halos (tens of parsecs). As a result, the $\gamma$-ray surface brightness ($\gamma$SB) distribution of pulsar halos provides an accurate projection of their parent electron spatial distribution, making them ideal probes for investigating cosmic-ray (CR) propagation in the localized ISM. Furthermore, the $\gamma$-ray spectral energy distribution of pulsar halos is determined by the energy spectrum of escaped electrons, offering a unique opportunity to study electron escape processes from PWNe \cite{Fang:2022mdg}.

Pulsar halos primarily demonstrate the slow-diffusion behavior of CR electrons in the ISM surrounding pulsars, with derived diffusion coefficients two orders of magnitude lower than the Galactic average \cite{Abeysekara:2017old}. This suppressed diffusion enables sufficient electron accumulation around pulsars, making these halos detectable. Nearly ten pulsar halos or candidates have been detected to date \cite{Abeysekara:2017old,HAWC:2024scl,Aharonian:2021jtz,Fang:2022qaf,LHAASO:2024flo,HAWC:2023jsq,Albert:2020fua,2025arXiv250801934K}, with ongoing observations continually expanding the sample. Both direct and indirect evidence suggest that pulsar halos could be a common phenomenon \cite{DiMauro:2019hwn,Albert:2025gwm,2025arXiv250317442J}, though not necessarily universal \cite{Fang:2019ayz,Martin:2022hrx,2025arXiv250701495B}.

The origin of slow diffusion has sparked extensive discussion, yet no consensus has been reached \cite{Evoli:2018aza,Kun:2019sks,Liu:2019zyj,Recchia:2021kty,Mukhopadhyay:2021dyh,DeLaTorreLuque:2022chz}. Determining the origin of slow diffusion is crucial for understanding the magnetohydrodynamic (MHD) characteristics of the localized ISM, and would significantly impact the interpretation of several key issues including the positron excess (e.g., \cite{Hooper:2017gtd,Fang:2018qco,Profumo:2018fmz,Manconi:2020ipm,Schroer:2023aoh}) and diffuse $\gamma$-ray excess (e.g., \cite{Linden:2017blp,Dekker:2023six,Yan:2023hpt}). Future measurements of energy-dependent $\gamma$SB distributions with improved sensitivity and spatial resolution could provide critical constraints for distinguishing between these competing hypotheses.

During the initial discovery stage of pulsar halos, their $\gamma$SB distributions were adequately explained by a homogeneous, isotropic electron diffusion-loss model, termed here as the standard diffusion model. Within this framework, the $\gamma$SB profile could also be approximated by simplified formulas \cite{Abeysekara:2017old,Aharonian:2021jtz}.
However, given the diverse possible origins of slow-diffusion phenomena, the actual electron transport processes are expected to be significantly more complex. With improving observational precision, the standard diffusion model or the simplified formulas are expected to become insufficient for interpreting upcoming measurements\footnote{Moreover, using these over-simplified formulas for estimating physical parameters may lead to incorrect conclusions. \cite{Guo:2024uuf}.}.

Given that models beyond standard diffusion involve computationally intricate procedures, this work introduces a lightweight software for pulsar halo emission called the Pulsar Halo Emission Computation Tool (\texttt{PHECT}), developed in \texttt{C}. The tool requires minimal dependencies (\texttt{GSL} and \texttt{YAML}), and users can perform essential computations by merely configuring a \texttt{YAML} file. \texttt{PHECT} provides a variety of physical and phenomenological models of pulsar halos for selection, which will accommodate the growing number of pulsar halo samples and increasingly precise observations. The software is available at \url{https://code.ihep.ac.cn/fangkun/phect}.

This paper is organized as follows. Section~\ref{sec:model} briefly reviews the mechanism of pulsar halos and possible interpretations for the slow-diffusion phenomenon. Section~\ref{sec:method} introduces the computational methods of different models along with the parameter configuration. Section~\ref{sec:output} presents the basic outcomes of \texttt{PHECT}. Section~\ref{sec:test} verifies the reliability of the computational results. Section~\ref{sec:conclu} provides concluding remarks and prospects for the tool.

\section{Origin of pulsar halos}
\label{sec:model}
Young pulsars typically power a PWN located at the center of their host supernova remnants (SNRs), as their velocities are significantly lower than the expansion speeds of young SNRs. As the expansion decelerates over time, the pulsar becomes offset from its initial position, gradually departing from the original PWN. When the pulsar approaches the SNR boundary or escapes the remnant, its motion exceeds the local sound speed, causing a newly formed bow-shock PWN. The size of the bow-shock PWN is constrained by the ram pressure from the motion of the pulsar, typically not exceeding $\simeq1$~pc \cite{Gaensler:2006ua}.

Electrons that escape from the bow-shock PWN freely diffuse into the surrounding medium. Through ICS with background photons, they generate $\gamma$-ray emission that forms a pulsar halo\footnote{There are also discussions about the possibility of pulsar halos being of hadronic origin \cite{Yang:2021iwe}.}. A defining characteristic of the halo is the significantly larger spatial extent compared to the associated bow-shock PWN \cite{Giacinti:2019nbu,Fang:2022fof}, as it originates from freely diffusing electrons. Additionally, pulsar halo searches typically focus on middle-aged pulsars ($\gtrsim50$~kyr), ensuring sufficient spatial separation from the relic of the initial PWN.

Despite the extended nature of pulsar halos (tens of parsecs), the derived diffusion coefficient is significantly smaller than the Galactic average suggested by the secondary-to-primary flux ratios of CR nuclei (e.g., \cite{Yuan:2017ozr}). The origin of this slow-diffusion phenomenon is the primary enigma in understanding the formation mechanism of pulsar halos. The differences among models in \texttt{PHECT} partially stem from their distinct assumptions about the slow-diffusion origin. 

\begin{figure}[!htb]
\centering
\includegraphics[width=0.48\textwidth]{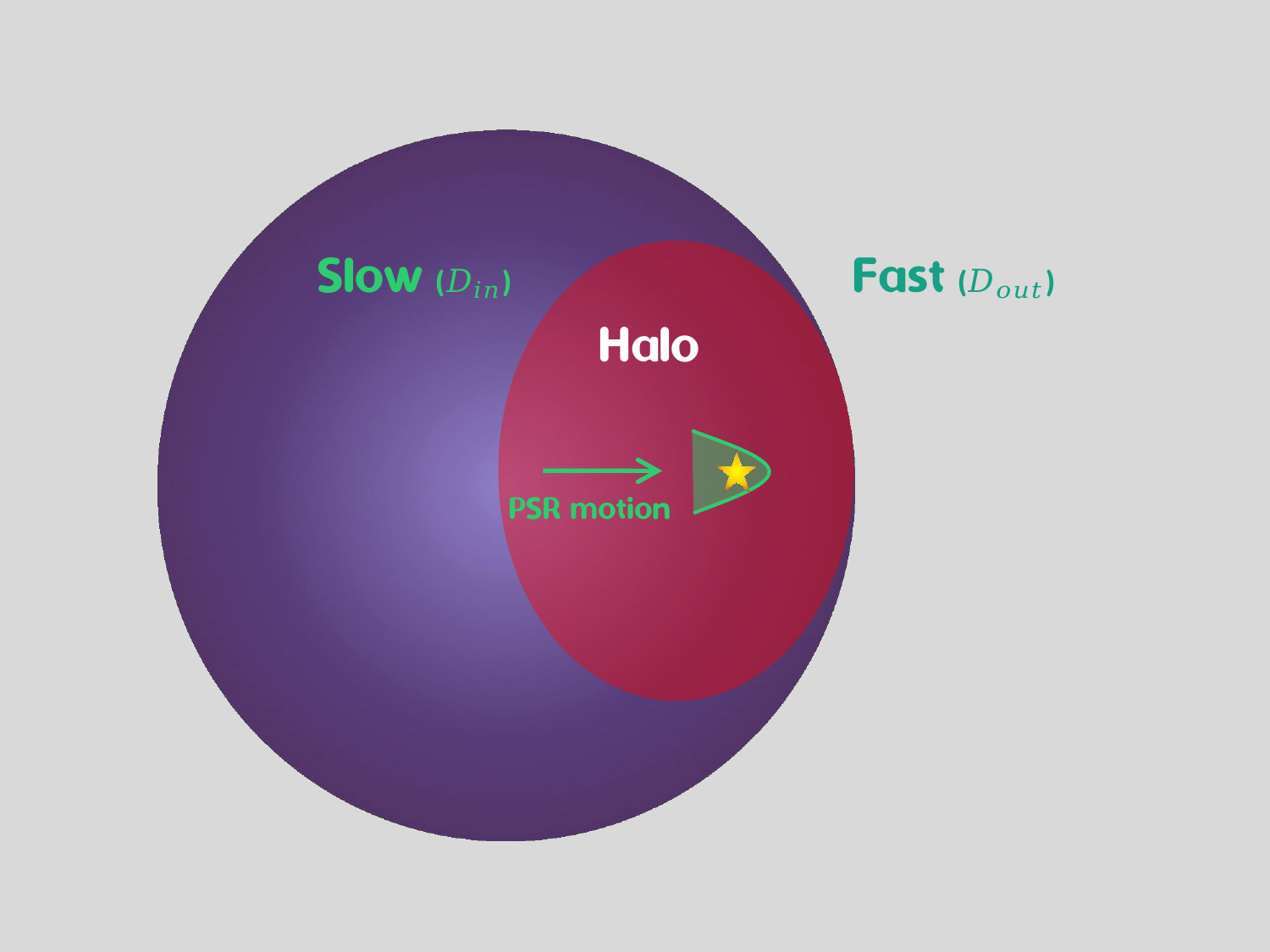}
\includegraphics[width=0.48\textwidth]{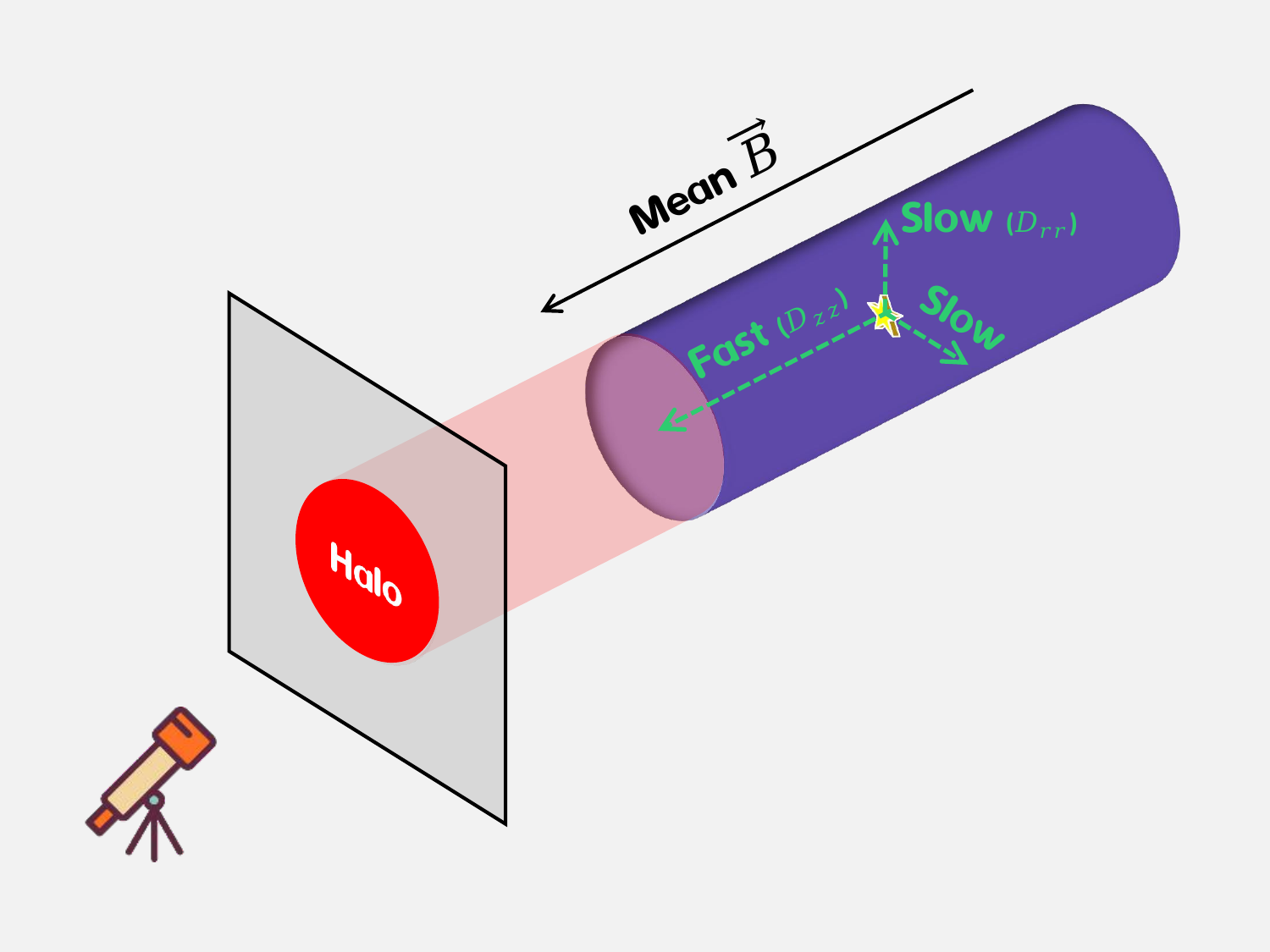}
\caption{Schematic illustrations of two of the possible slow-diffusion interpretations. The star symbols denote the current position of the pulsar. Left: SNR-induced model. The slow diffusion originates from a strong turbulence region (blue area) generated by the associated SNR or the progenitor stellar wind of the pulsar. The small green parabolic zone near the pulsar schematically represents the bow-shock PWN. $D_\mathrm{in}$ and $D_\mathrm{out}$ are the diffusion coefficients inside and outside the SNR, respectively. Right: Anisotropic diffusion model. The apparent slow diffusion results from the projection effects of anisotropic diffusion. The blue region illustrates the asymmetric electron distribution due to anisotropic diffusion conditions. $D_{zz}$ and $D_{rr}$ are the diffusion coefficients parallel and perpendicular to the mean magnetic field around the pulsar, respectively.}
\label{fig:sketch}
\end{figure}

According to resonant scattering theory, a lower CR diffusion coefficient implies stronger turbulent magnetic fields at the resonant scale. For pulsars, their host SNRs serve as a direct source of the required turbulent environment. A fraction of the SNR shock energy converts into turbulent energy, capable of suppressing the diffusion coefficient by two orders of magnitude within the remnant \cite{Kun:2019sks,2022ApJ...932...65W}. Consequently, pulsars still embedded within their SNRs naturally reside in such slow-diffusion environments \cite{Kun:2019sks}, as illustrated in the left panel of Fig.~\ref{fig:sketch}. Furthermore, the massive-star progenitors of pulsars can create low-density environments by blowing powerful winds \cite{2025arXiv250701495B}, resulting in a larger SNR size than typically expected, thereby increasing the likelihood of pulsar containment\footnote{The wind bubble itself could also be the required turbulent environment \cite{2025arXiv250701495B}.}. Among known pulsar halos, the central pulsar of the Monogem halo definitively resides within its host SNR \cite{Yao:2022cse}.
We refer to the above scenario as the SNR-induced model, which has been incorporated into \texttt{PHECT}.

An alternative mechanism for generating magnetic turbulence is the resonant streaming instability self-excited by escaping electrons \cite{Evoli:2018aza}. Neglecting the pulsar proper motion, this self-excited picture can reproduce the observed slow diffusion for certain parameter ranges \cite{Mukhopadhyay:2021dyh}. However, when accounting for pulsar motion, only recently escaped electrons contribute to turbulence generation, which may be insufficient to significantly reduce the diffusion coefficient \cite{Kun:2019sks}. The corresponding computation requires coupled solutions to both particle transport and magnetic turbulence evolution. The current version of \texttt{PHECT} does not include this mechanism, but we intend to integrate it in future updates.

It is also proposed that interpreting pulsar halo morphology may not require amplified magnetic turbulence \cite{Liu:2019zyj}. If the magnetic field correlation length near the pulsar is $\sim100$~pc, electron diffusion on the scale of pulsar halos exhibits notable anisotropy. In this scenario, the diffusion coefficient perpendicular to the magnetic field becomes much smaller than the parallel one, with the latter remaining comparable to the Galactic average. When the local magnetic field around the pulsar aligns closely with the line of sight (LOS) of the observer, the observed halo morphology becomes dominated by perpendicular diffusion. This results in the appearance of slow diffusion due to projection effects, as illustrated in the right panel of Fig.~\ref{fig:sketch}. This scenario sets strict requirements on the mean magnetic field orientation near pulsars, consequently predicting fewer observable pulsar halos \cite{DeLaTorreLuque:2022chz}. This anisotropic diffusion model has been incorporated into \texttt{PHECT}.

Most studies on CR propagation using diffusion equations typically neglect the issue of superluminal propagation. However, this issue becomes non-negligible for timescales $\ll D/c^2$. It is proposed that even the Galactic-average diffusion coefficient can produce a sharp $\gamma$SB near pulsars when a superluminal correction is considered \cite{Recchia:2021kty}. 
\texttt{PHECT} includes a superluminal-corrected version of the standard diffusion model \cite{Bao:2021hey}. Note that the correction is only significant for fast-diffusion scenarios. 

Recent MHD simulations suggest the possible existence of magnetic mirror points in the ISM, which could trigger mirror diffusion phenomena \cite{2024arXiv240512146B,2025arXiv250615031X}. The interplay between mirror diffusion and resonant scattering may produce slow-diffusion effects even without additional turbulence sources. A distinctive feature of this scenario is that particle transport during the initial escape phase follows superdiffusion behavior. \texttt{PHECT} currently includes a superdiffusive transport model for electrons. However, in real cases, the electron propagation may transition to standard diffusion as the escape time increases. We plan to take this feature into account in future versions.

\section{Computational methods}
\label{sec:method}
The primary objective of \texttt{PHECT} is to compute the $\gamma$SB of pulsar halos. The tool first solves the electron transport equation to derive the electron number density distribution. As a byproduct, this distribution can estimate the positron/electron flux at Earth originating from nearby pulsars. Subsequently, the LOS integration of electron densities yields the electron surface density distribution. Finally, combining this surface density with background photon fields, the $\gamma$SB is computed through ICS calculations. Unless otherwise specified, numerical integrations involved in the procedures are performed using the CQUAD algorithm in GSL.

All adjustable parameters in \texttt{PHECT} are configured via the \texttt{param\_config.yaml} file, with a comprehensive listing provided in Appendix~\ref{app:params}. Throughout this paper, such parameters are displayed in \texttt{monospaced font} for clarity.

\subsection{Electron propagation}
\label{subsec:electron}

\subsubsection{Common framework}
\label{subsubsec:common}
The electron propagation in the ISM is typically described by the diffusion-loss equation, written as:
\begin{equation}
 \frac{\partial N}{\partial t} = \nabla\cdot(D\nabla N) + \frac{\partial (bN)}{\partial E_e} + Q\,,
 \label{eq:prop}
\end{equation}
where $E_e$ is the electron kinetic energy, $N=N(E_e, \mathbi{r}, t)$ is the energy differential number density of electrons, $b=b(E_e)\equiv|dE_e/dt|$ is the absolute energy-loss rate, $Q=Q(E_e, \mathbi{r}, t)$ is the source function, and $D=D(E_e, \mathbi{r})$ is the diffusion coefficient\footnote{$D$ is a tensor under anisotropic diffusion.}.

For electrons with energies $E_e \gtrsim 1$~GeV, the primary energy-loss processes during propagation are synchrotron radiation in magnetic fields and ICS with background photons. The energy-loss rate can be expressed as $b=b_0E_e^2=(b_{0,\mathrm{syc}}+b_{0,\mathrm{ics}})E_e^2$, and the factor of the synchrotron term is (as given in, e.g., Ref~\cite{Blumenthal:1970gc})
\begin{equation}
 b_{0,\mathrm{syc}} = \frac{\sigma_TcB^2}{6\pi(m_ec^2)^2}\,,
 \label{eq:loss_syn}
\end{equation}
where $B$ is the magnetic field strength, which is set as an adjustable parameter, \texttt{B}. In the current version, $B$ is assumed to be uniform throughout the space. If the actual $B$ exhibits non-negligible variation within the halo region, it could impact the determination of the diffusion coefficient. Detecting the x-ray emission from pulsar halos in the future may provide a way to estimate the spatial pattern of $B$ through a multiwavelength analysis. In Eq.~(\ref{eq:loss_syn}), $c$ is the speed of light, $\sigma_T$ is the Thomson scattering cross section, and $m_e$ is the electron rest mass.

Taking the Klein-Nishina effect into account, the factor of the ICS term can be written as
\begin{equation}
 b_{0,\mathrm{ics}} = \sum_i \frac{20\sigma_Tcw_i}{\pi^4(m_ec^2)^2}Y(E_e, T_i)\,,
 \label{eq:loss_ics}
\end{equation}
where $T_i$ and $w_i$ are the temperature and energy density of the background photon fields, respectively. The components are assumed to be the cosmic microwave background (CMB), the infrared radiation from interstellar dust, and the starlight. Among these, the CMB spectrum takes a blackbody form, while the dust radiation and starlight radiation are assumed to follow a graybody spectrum. The temperatures and energy densities of these components are named after (\texttt{T\_cmb}, \texttt{density\_cmb}), (\texttt{T\_dust}, \texttt{density\_dust}), and (\texttt{T\_sl}, \texttt{density\_sl}) in the configuration file, respectively. When the background photon spectrum is either blackbody or graybody, the variable in the function $Y$ can be simplified to $x\equiv 4E_ek_BT_i/(m_ec^2)^2$. The form of $Y(x)$ can be sufficiently approximated using polynomial interpolation \cite{Fang:2020dmi}.

As introduced in Sec.~\ref{sec:model}, PWNe can be regarded as the electron source for pulsar halos. We decompose the source function $Q$ into a spatial component $q_r$, a temporal component $q_t$, and an energy component $q_E$. As the bow-shock PWNe are much smaller than the halos, their spatial distribution can be safely assumed to be a $\delta$ function as
\begin{equation}
 q_r(\mathbi{r})=\delta(\mathbi{r})\,,
 \label{eq:src_r}
\end{equation}
where the spatial origin is set at the pulsar position.

The temporal variation of electron injection from PWNe is assumed to follow that of the pulsar spin-down luminosity, as
\begin{equation}
 q_t(t)=\left\{
 \begin{aligned}
  & (1+t/\tau_0)^{-2}/(1+t_s/\tau_0)^{-2}\,, \quad & t\geq0 \\
  & 0\,, \quad & t<0 \\
 \end{aligned}
 \right.\,,
 \label{eq:src_t}
\end{equation}
where $t=0$ represents the pulsar birth time, $t_s$ is the pulsar current age, and $\tau_0$ is the initial spin-down timescale (referred to as \texttt{tau0} in the configuration file), with the assumption that the pulsar braking index is $n=3$. Few pulsars have an estimated $\tau_0$ \cite{Migliazzo:2002vd,Popov:2012ng,Suzuki:2021ium}. For the Crab pulsar, $\tau_0 \approx 950$ yr, whereas for PSR B1951$+$32, $\tau_0 \approx 40$ kyr \cite{Migliazzo:2002vd}, indicating that there could be considerable fluctuation in $\tau_0$. However, when $t_s\gg\tau_0$, the uncertainty in $\tau_0$ has little impact on the TeV $\gamma$SB of pulsar halos. Additionally, the pulsar catalog \cite{Manchester:2004bp} offers the pulsar characteristic age, which is different from $t_s$. For convenience, we use the characteristic age, denoted as \texttt{ts\_c}, in the configuration file, which relates to the true age by $t_s = \mathtt{ts\_c} - \tau_0$. Therefore, \texttt{tau0} should be set to be less than \texttt{ts\_c}.

In accordance with Eq.~(\ref{eq:src_t}), $q_E$ represents the electron injection spectrum at the current time, that is, 
\begin{equation}
 q_E(E_e)=q_{E,0}\,E_e^{-p}\,{\rm exp}\left[-\left(\frac{E_e}{E_{e,c}}\right)^s\right]\,,
 \label{eq:src_e}
\end{equation}
the form of which is suggested by the relativistic shock acceleration theory \cite{Dempsey:2007ng}. Assuming the conversion efficiency, $\eta$, from the pulsar spin-down energy to the energy of escaping electrons is time-independent, we have the following relation:
\begin{equation}
 \eta\dot{E}_{\mathrm{rot}}=\int q_E(E_e)E_edE_e\,,
 \label{eq:eta}
\end{equation}
where $\dot{E}_{\mathrm{rot}}$ is the current spin-down luminosity of pulsars, and we adopt $1$~GeV and $50E_{e,c}$ as the lower and upper limits\footnote{In models using numerical solutions, the upper limit is set as \texttt{Ee\_max} rather than $50E_c$. Please refer to Appendix~\ref{app:numr} for details.} for the integral, respectively. 

Currently, there is no robust constraint on the lowest energy of the injection spectrum, which may affect the estimation of $\eta$. However, spectral measurements of pulsar halos and PWNe generally support $p<2$ \cite{Shao-Qiang:2018zla,Fang:2021qon,Fang:2022qaf,HAWC:2024scl,Reynolds:2017hbs}. In this scenario, the total energy of the injection spectrum is concentrated at the high-energy end \cite{Malyshev:2009tw}, which means that the uncertainty in the lowest energy has almost no impact on the computed results.

We choose $\eta$ as an adjustable parameter instead of $q_{E,0}$ because $\eta$ is more physically intuitive. In the configuration file, $\dot{E}_{\mathrm{rot}}$, $p$, $s$, $E_{e,c}$, and $\eta$ correspond to \texttt{Edot\_now}, \texttt{index\_p}, \texttt{index\_exp}, \texttt{Ec}, and \texttt{eta}, respectively. Among them, \texttt{Edot\_now} can be obtained from the pulsar catalog \cite{Manchester:2004bp}, and \texttt{index\_exp} is set to $2$ as suggested by Ref.~\cite{Dempsey:2007ng}.

\subsubsection{Model-specific modules}
\label{subsubsec:indiv}
The differences between the models in \texttt{PHECT} primarily manifest in the procedures for solving the electron propagation equation. In addition to the physically motivated models discussed in Sec.~\ref{sec:model}, \texttt{PHECT} also includes phenomenological models such as standard diffusion and spherically symmetric two-zone diffusion. These models are valuable for estimating basic physical parameters and performing statistical analyses of pulsar halos. All models in \texttt{PHECT} are categorized into two principal classes: spherically symmetric models and cylindrically symmetric models. The model selection is controlled by the \texttt{model} parameter. The time required for a full run for each model is summarized in Appendix~\ref{app:time}.

For a spherically symmetric model, the electron number density is stored in a 2D array of size (\texttt{NUM\_Ee}, \texttt{NUM\_r}). The electron energy is logarithmically divided into \texttt{NUM\_Ee} grid points within the range of \texttt{Ee\_max} and \texttt{Ee\_min}. For the spatial dimension grid points, we use a tangent form \cite{Porter:2021tlr}, where the value of the $j$-th grid point is given by
\begin{equation}
 r[j] = \frac{dr}{a}\tan(aj)\,,
 \label{eq:grid_r}
\end{equation}
where
\begin{equation}
 \begin{aligned}
  & a = \frac{1}{\mathtt{NUM\_r}-1}\,\arctan\left\{\frac{\mathtt{r\_max}}{\mathtt{r\_ref}}\tan\left[\arccos\left(\frac{1}{\sqrt{\mathtt{r\_ampl}}}\right)\right]\right\}\,, \\
 & dr = a\,\frac{\mathtt{r\_ref}}{\tan\left[\arccos\left(\frac{1}{\sqrt{\mathtt{r\_ampl}}}\right)\right]}\,. \\
 \end{aligned}
 \label{eq:grid_r2}
\end{equation}
Equation (\ref{eq:grid_r}) features linearly spaced points near the source with a step size of $\approx dr$, with gradually expanding spacing at larger distances. At \texttt{r\_ref}, the step size increases by $\texttt{r\_ampl}\times dr$. When comparing the model to the $\gamma$SB measurements of the pulsar halo, our primary focus is on $N$ within several tens of parsecs around the pulsar. Therefore, this tangent grid effectively reduces spatial grid requirements, significantly boosting computational efficiency. However, when estimating the electron energy spectrum \textit{reaching Earth} from a pulsar, accuracy at large distances from the pulsar becomes crucial. In this case, we recommend either increasing \texttt{r\_ref} or setting \texttt{r\_ref} equal to \texttt{r\_max}, thereby reducing to a linear grid.

It should be noted that certain models within \texttt{PHECT} employ numerical methods, where the spatial discretization is based on Eq.~(\ref{eq:grid_r}), but the specifics differ slightly. For further details, please refer to Appendix \ref{app:numr}.

The following introduces the spherically symmetric models. We use the suffixes `A' and `N' to distinguish whether the model employs a semi-analytical or a numerical solution.

\begin{itemize}
 \item \texttt{StdDiff\_A}:
 This is the standard diffusion model, the most basic one among \texttt{PHECT}. When interpreting the initial measurements of the Geminga and Monogem halos \cite{Abeysekara:2017old}, standard diffusion achieves an adequate fit, indicating that describing the electron propagation within pulsar halos by a diffusion process is generally appropriate. The diffusion coefficient takes the form of
   \begin{equation}
    D(E_e)=D_0\left(\frac{E_e}{100~\mathrm{TeV}}\right)^\delta\,,
    \label{eq:d_norm}
   \end{equation}
 where $D_0$ and $\delta$ are referred to as \texttt{D0} and \texttt{delta} in the configuration file, respectively. In this case, Eq.(\ref{eq:prop}) can be solved using the Green's function method \cite{1964ocr..book.....G}. The form of the solution is detailed in Appendix \ref{app:ana}. Numerical integrations within the Green's function are rigorously managed. 
 \item \texttt{StdDiff\_A0}:
 This model is identical to the \texttt{StdDiff\_A} model, while approximations are employed for terms requiring numerical integrations within the Green's function, as explained in Appendix \ref{app:ana}. It is recommended only for cases where computational resources are limited or when a rough estimate is sufficient.
 \item \texttt{StdDiff\_N}:
 This model also describes standard diffusion, but employs a numerical scheme to solve Eq.~(\ref{eq:prop}). It can be viewed as a special case of the \texttt{2Zone\_N} model (detailed later), where the two‑zone configuration reduces to a single‑zone diffusion. As shown in Sec.~\ref{sec:test}, this model produces results consistent with \texttt{StdDiff\_A} while achieving significantly higher computational speed. For these reasons, it has been designated the default model in PHECT.
 \item \texttt{Juttner\_A}:
 As outlined in Sec.~\ref{sec:model}, the diffusion equation has the superluminal issue. This model addresses the superluminal problem by reformulating the Green's function as a J\"{u}ttner propagator \cite{Aloisio:2008tx} (hence the ``J'' designation). While this approach—analogous to relativistic corrections of the Maxwell-Boltzmann distribution—lacks full rigor, it yields particle distributions that closely approximate recent rigorous solutions after time integration \cite{Lv:2024bxo,Kawanaka:2024qiu}. For pulsar halos where the electron injection is continuous, the approximation error is negligible \cite{Lv:2024bxo}. Considering computational efficiency, we still use this approximate method. The form of the solution is provided in Appendix \ref{app:ana}.
 \item \texttt{Super\_A}:
 This model describes electron propagation by superdiffusion, replacing the standard Laplace operator in Eq.~(\ref{eq:prop}) with a fractional Laplace operator $\Delta^{\alpha/2}$. The superdiffusion index \texttt{alpha} determines the transport regime; when $\alpha=2$, the system reduces to standard diffusion. Although the form of the diffusion coefficient under this model is the same as in Eq.~(\ref{eq:d_norm}), both its dimensions and meaning differ. The form of the solution is provided in Appendix \ref{app:ana}.
 \item \texttt{2Zone\_N}:
 This is a two-zone diffusion model, solved using a numerical method. The diffusion coefficient takes the form of
 \begin{equation}
  D(E_e, r)=\left\{
  \begin{aligned}
   & D_\mathrm{in}\,, \quad & r\leq \mathtt{r\_2z} \\
   & D_\mathrm{out}\,, & \quad r>\mathtt{r\_2z} \\
  \end{aligned}
  \right.\,,
  \label{eq:d_2z}
 \end{equation}
 where $D_\mathrm{in}=D_0(E_e/100~\mathrm{TeV})^\delta$, $D_\mathrm{out}=\mathtt{ratio\_2z}\cdot D_\mathrm{in}$, and $\mathtt{ratio\_2z}\gg1$. The slow-diffusion phenomenon is unlikely to be universal throughout the Galaxy \cite{Hooper:2017gtd,Hooper:2017tkg}. Whether the slow-diffusion zone is SNR-induced or self-excited, it exists only near pulsars, and the diffusion pattern can be considered a two-zone-like form. Although the distribution of diffusion coefficients in these scenarios could be more complex than Eq.~(\ref{eq:d_2z}), this phenomenological model can effectively capture certain characteristics that distinguish it from one-zone slow diffusion. These include, e.g., the impact on the spectral shape of $\gamma$-ray emission or on the electron/positron flux reaching Earth from pulsars \cite{Fang:2023xla}.

 We employ a numerical approach for solving the two-zone diffusion problem \cite{Fang:2018qco}. Unlike CR propagation codes such as \texttt{GALPROP} \cite {Porter:2021tlr} and \texttt{DRAGON} \cite{Evoli:2016xgn}, we use the finite volume method to discretize the diffusion equation. This approach enforces electron flux conservation even when handling non-uniform grids (e.g., Eq.~(\ref{eq:grid_r})) and discontinuous diffusion coefficients (e.g., Eq.~(\ref{eq:d_2z})). The complete mathematical derivation of our discretization scheme is presented in Appendix~\ref{app:numr}.
\end{itemize}

For cylindrically symmetric models, the electron number density is stored in a 3D array of size (\texttt{NUM\_Ee}, \texttt{NUM\_r}, \texttt{2NUM\_z-1}), where \texttt{NUM\_z} is the number of grid points from $z=0$ to $z=\mathtt{z\_max}$. The grid setting along the $z$-axis is similar to the tangent function described in Eq.~(\ref{eq:grid_r}). The angle between the $z$-axis and the vector from the source to the observer is referred to as \texttt{PHI}. The following introduces the cylindrically symmetric models.

\begin{itemize}
 \item \texttt{SNR2Z\_N}:
 This is the SNR-induced model introduced in Sec.~\ref{sec:model}. Under realistic conditions, the diffusion coefficient within SNRs may vary spatially \cite{2022ApJ...932...65W}. Nevertheless, for simplicity, we assume a constant value inside the SNR, resulting in a two-zone diffusion pattern (hence the ``2Z'' designation):
 \begin{equation}
  D(E_e, r, z)=\left\{
  \begin{aligned}
   & D_\mathrm{in}\,, \quad & R< \mathtt{R\_snr} \\
   & D_\mathrm{out}\,, & \quad R>\mathtt{R\_snr} \\
  \end{aligned}
  \right.\,,
  \label{eq:d_snr}
 \end{equation}
 where $D_\mathrm{in}=D_0(E_e/100~\mathrm{TeV})^\delta$, $D_\mathrm{out}=\mathtt{ratio\_snr}\cdot D_\mathrm{in}$, and $\mathtt{ratio\_snr}\gg1$. $R$ is the distance to the SNR center, defined as $R^2=(r-\mathtt{r\_snr})^2+(z-\mathtt{z\_snr})^2$, where (\texttt{r\_snr}, \texttt{z\_snr}) is the coordinates of the SNR center. Phenomenologically, the primary difference between this model and the \texttt{2Zone\_N} model is the possible asymmetry of the pulsar position relative to the slow-diffusion zone. The $z$-axis orientation is chosen as the direction of the pulsar motion. When \texttt{PHI} is $0^\circ$ or $180^\circ$, the morphology of $\gamma$SB is symmetrical. However, when \texttt{PHI} approaches $90^\circ$ or $270^\circ$, the pulsar halo may exhibit significant asymmetry. We employ a numerical method to solve this model, and the derivation of the discretization scheme is presented in Appendix~\ref{app:numr}.
 \item \texttt{Aniso\_A}:
 This is the anisotropic diffusion model introduced in Sec.~\ref{sec:model}. In this scenario, the diffusion term in Eq.~(\ref{eq:prop}) is expressed as
 \begin{equation}
  \frac{D_{rr}}{r}\frac{\partial}{\partial r}\left(r\frac{\partial N}{\partial r}\right)+D_{zz}\frac{\partial^2N}{\partial z^2}\,,
  \label{eq:diff_aniso}
 \end{equation}
 where $D_{rr}=D_0(E_e/100~\mathrm{TeV})^\delta$, $D_{zz}=M_A^{-4}D_{rr}$, and $M_A$ is the Alfv\'{e}nic Mach number, referred to as \texttt{Ma} in the configuration file. In the anisotropic diffusion scenario, $D_{zz}$ is required to be consistent with the Galactic average value, which means that \texttt{Ma} is approximately $0.1-0.2$. The $z$-axis orientation is defined by the local mean magnetic field direction near the pulsar. When the $z$-axis deviates from the LOS (i.e., $\mathtt{PHI}\neq0$), the pulsar halo exhibits asymmetric morphology. Assuming $D_{rr}$ and $D_{zz}$ are spatially independent, coordinate transformation enables a semi-analytical solution of this model \cite{Fang:2023axu}. The form of the solution is provided in Appendix \ref{app:ana}.
 \item \texttt{Aniso\_N}: This model also describes anisotropic diffusion but employs a numerical method, achieving significantly higher computational speed compared to \texttt{Aniso\_A}. The discretization scheme is presented in Appendix~\ref{app:numr}. 
\end{itemize}

\subsection{\texorpdfstring{$\gamma$-ray emission}{gamma-ray emission}}
\label{subsec:gamma}
We first perform the LOS integration on the electron number density solved in Sec.~\ref{subsec:electron} to derive the surface distribution $S_e$.

For the spherically symmetric cases, we have 
\begin{equation}
 S_e(E_e,\theta)=\int_0^\infty N(E_e, r)dl\,,
 \label{eq:s_sph}
\end{equation}
where $\theta$ is the angle between the LOS and the direction from the observer to the pulsar, and $l$ is the distance from a point along the LOS to the observer. Denoting the distance between the pulsar and the Earth as \texttt{rs}, the relation between $l$ and $r$ is $r=\sqrt{\mathtt{rs}^2+l^2-2\mathtt{rs}\,l\cos\theta}$. We perform numerical integration of Eq.~(\ref{eq:s_sph}), employing linear interpolation to determine $N$ at specific spatial coordinates. Given the rapid decline in $N$ with distance from the pulsar, the integration domain need not extend to infinity. A LOS bound scale \texttt{los\_bound} is set to restrict the integration within this radial distance from the pulsar. The integration limits can then be derived as
\begin{equation}
 \begin{aligned}
  & l_{\max} = \mathtt{rs}\,\cos\theta+\sqrt{\mathtt{los\_bound}^2-(\mathtt{rs}\,\sin\theta)^2}\,,\\
  & l_{\min} = \mathtt{rs}\,\cos\theta-\sqrt{\mathtt{los\_bound}^2-(\mathtt{rs}\,\sin\theta)^2}\,.\\
 \end{aligned}
 \label{eq:los_bound}
\end{equation}

After obtaining $S_e$, we can further determine the $\gamma$SB generated by ICS between the electrons and background photons:
\begin{equation}
 s_\gamma(E_\gamma,\theta)=\frac{1}{4\pi}\iint S_e(E_e,\theta)f(E_e,\epsilon,E_\gamma)dE_ed\epsilon\,,
 \label{eq:sg}
\end{equation}
where $E_\gamma$ is the energy of emitted $\gamma$-ray photons, $\epsilon$ is the energy of background photons, and $f$ is the production rate of $\gamma$-ray photons resulting from the scattering of a single electron with a background photon. The background photon field configuration is identical to that used for computing electron energy losses (Sec.~\ref{subsubsec:common}). Equation (\ref{eq:sg}) assumes isotropic ICS, whereas the \texttt{Juttner\_A} model considers a quasi-ballistic propagation stage for initially escaped electrons, thus requiring a correction \cite{Recchia:2021kty}. We have equivalently integrated this correction into the $S_e$ calculation, enabling consistent ICS computation across all models.

\texttt{PHECT} stores $s_\gamma$ in a 2D array of size (\texttt{NUM\_Eg}, \texttt{NUM\_tht}) for spherically symmetric models. The $\gamma$-ray energy grid is logarithmically spaced between \texttt{Eg\_min} and \texttt{Eg\_max}, while the angular grid follows a tangent form analogous to Eq.~(\ref{eq:grid_r}). When comparing the model to observation data, we often need to compute the $\gamma$SB profile within a certain energy range:
\begin{equation}
 S_\gamma(\theta)=
 \int s_\gamma(E_\gamma,\theta)E_\gamma dE_\gamma\,,
 \label{eq:Sg}
\end{equation}
or the energy spectrum within a certain angular range:
\begin{equation}
 F(E_\gamma)=
 \int s_\gamma(E_\gamma,\theta)2\pi\theta d\theta\,.
 \label{eq:spec}
\end{equation}
The tool employs trapezoidal integration for Eqs.~(\ref{eq:Sg}) and (\ref{eq:spec}), and users can adjust grid densities for precision control.

Sometimes, experiments provide $\gamma$SB in terms of photon number instead of energy \cite{HAWC:2024scl}, that is,
\begin{equation}
 S_\gamma(\theta)=
 \int s_\gamma(E_\gamma,\theta) dE_\gamma\,.
 \label{eq:Sg2}
\end{equation}
In the configuration file, the variable \texttt{gunit} is used to control the units of $\gamma$SB. When \texttt{gunit} is set to \texttt{energy} (\texttt{counts}), the definition of $\gamma$SB follows Eq.~(\ref{eq:Sg}) (Eq.~(\ref{eq:Sg2})). 

As pulsar halo measurements achieve higher precision, convolving $S_\gamma$ with the point spread function (PSF) becomes necessary for the model-data comparison, even when the source extent far exceeds the PSF size. \texttt{PHECT} assumes a symmetric 2D Gaussian PSF with width $\sigma_\mathrm{PSF}$, producing convolved $\gamma$SB:
\begin{equation}
 \widetilde{S}_\gamma(\theta) = (S_\gamma*g)(\theta)=\int_0^\infty\int_0^{2\pi}S_\gamma(\theta')g(\sqrt{\theta^2+\theta'^2-2\theta\theta'\cos\varphi})\theta'd\varphi d\theta'\,,
 \label{eq:psf}
\end{equation}
where $g$ is the PSF. While the convolution is intrinsically two-dimensional, the adopted Gaussian PSF form enables dimensionality reduction through the modified Bessel function, thereby requiring only 1D numerical integration.

For cylindrically symmetric models, the LOS integration on $N$ takes the form of
\begin{equation}
 S_e(E_e,\varphi,\theta)=\int_0^\infty N(E_e,r,z)dl\,,
 \label{eq:s_cylin}
\end{equation}
where the mapping from $(\varphi, \theta, l)$ to $(r, z)$ is referred to Ref.~\cite{Liu:2019zyj}, and can be expressed as
\begin{equation}
 \begin{aligned}
  & x=(\mathtt{rs}-l\cos\theta)\sin(\mathtt{PHI})
    +l\sin\theta\cos\varphi\cos(\mathtt{PHI})\,,\\
  & y=l\sin\theta\sin\varphi\,,\\
  & z=(\mathtt{rs}-l\cos\theta)\cos(\mathtt{PHI})
    -l\sin\theta\cos\varphi\sin(\mathtt{PHI})\,,\\
  & r=\sqrt{x^2+y^2}\,.\\
 \end{aligned}
 \label{eq:los_mapping}
\end{equation}
It should be noted that the azimuthal angle $\varphi$ in \texttt{PHECT} relates to the angle $\zeta$ in Fig.~1 of Ref.~\cite{Liu:2019zyj} by $\varphi=180^\circ-\zeta$. Furthermore, the symbol $\varphi$ denoting the inclination angle between $B$ and the $z$-axis in Fig.~1 of Ref.~\cite{Liu:2019zyj} corresponds to the parameter \texttt{PHI} in the configuration file of \texttt{PHECT}.

Analogous to the spherically symmetric cases, we can obtain $s_\gamma(E_\gamma,\varphi,\theta)$, $S_\gamma(\varphi,\theta)$, $\widetilde{S}_\gamma(\varphi,\theta)$, and $F(E_\gamma)$ for the cylindrically symmetric models, where $s_\gamma$ is stored in a 3D array of size (\texttt{NUM\_Eg}, $2\mathtt{NUM\_phi}+1$, \texttt{NUM\_tht}). Given the inherent mirror symmetry in cylindrically symmetric models, it is sufficient to compute $\gamma$SB for $\varphi\in[0, 180^\circ]$. The azimuthal grid is uniformly defined as $\varphi_i=i\,\Delta\varphi$ over the interval of $[0, 360^\circ]$, with $\Delta\varphi=180^\circ/\mathtt{NUM\_phi}$. Consequently, there are $2\mathtt{NUM\_phi}+1$ grid points in the azimuth dimension. 

\section{Output products}
\label{sec:output}
The default output files of \texttt{PHECT} include
\begin{itemize}
 \item Electron number density distribution ($N$): \texttt{[root]\_Ne.dat},
 \item 1D $\gamma$SB profile(s) ($\widetilde{S}_\gamma$): \texttt{[root]\_SBprof\_bin0.dat}, \texttt{[root]\_SBprof\_bin1.dat}, ... ,
 \item 2D $\gamma$SB map(s) ($\widetilde{S}_\gamma$): \texttt{[root]\_SBmap\_bin0.dat}, \texttt{[root]\_SBmap\_bin1.dat}, ... ,
 \item $\gamma$-ray energy spectrum ($F$): \texttt{[root]\_spec.dat},
\end{itemize}
where \texttt{root} is the user-assigned root name for the outputs. As the $\gamma$SB of spherically symmetric models does not depend on azimuth, \texttt{[root]\_SBmap\_bin*.dat} is exclusive for cylindrically symmetric models.

As noted in Sec.~\ref{sec:method}, while \texttt{[root]\_Ne.dat} is an intermediate product in pulsar halo computations, the derived quantity $c/(4\pi)N$ is meaningful for estimating local electron and positron fluxes from nearby pulsars ($\mathtt{rs}\lesssim1$~kpc). Even when different models produce comparable $\gamma$SB distributions for pulsar halos, their extrapolated electron and positron fluxes at Earth can exhibit substantial differences (e.g., \cite{Fang:2023xla,Xia:2024utu}).

For spherically symmetric models, the data stored in files named \texttt{[root]\_SBprof\_bin*.dat} represent the 1D $\gamma$SB profiles after the PSF convolution. The suffixes---\texttt{bin0}, \texttt{bin1}, etc.---denote different $\gamma$-ray energy bins, with the total number of bins determined by the parameter \texttt{NUM\_energy}. Each energy bin is defined by lower and upper bounds specified in the arrays \texttt{Eg\_1[NUM\_energy]} and \texttt{Eg\_2[NUM\_energy]}, respectively. Users may configure the PSF width individually for each energy bin, with these values stored in the array \texttt{sgm\_psf[NUM\_energy]}.

\begin{figure}[!htb]
\centering
\includegraphics[width=0.6\textwidth]{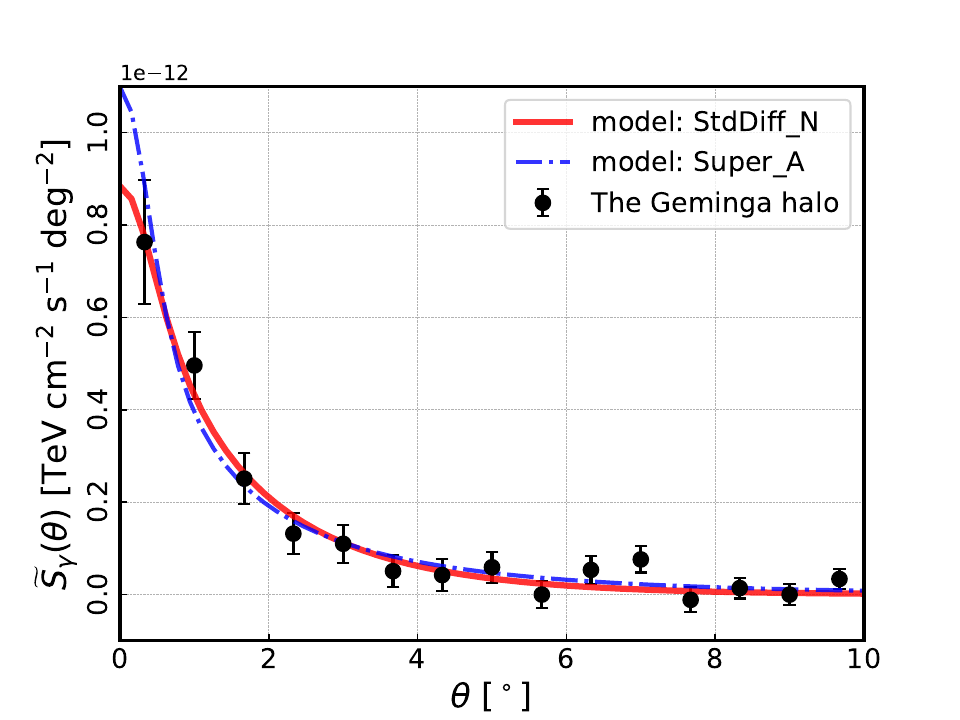}
\caption{One-dimensional $\gamma$SB of the Geminga halo computed by the standard diffusion (\texttt{StdDiff\_N}) and superdiffusion (\texttt{Super\_A}) models, in comparison with the HAWC data \cite{Abeysekara:2017old}. The parameters used are the default settings in \texttt{param\_config.yaml}, with the following exceptions: \texttt{D0=4.1e27}\footnote{Note that throughout this paper, parameters displayed in a monospaced font correspond to the unitless input values expected in \texttt{param\_config.yaml}. The corresponding units for each parameter are summarized in Table~\ref{tab:params}.} and \texttt{eta=0.049} for the \texttt{StdDiff\_N} model; \texttt{alpha=1.6}, \texttt{D0=1.5e20}, and \texttt{eta=0.070} for the \texttt{Super\_A} model.}
\label{fig:prof}
\end{figure}

Figure~\ref{fig:prof} presents the 1D $\gamma$SB of the Geminga halo in $8–40$~TeV computed from the standard diffusion model (\texttt{StdDiff\_N}) and the superdiffusion model (\texttt{Super\_A}) with \texttt{alpha=1.6}, in comparison with the HAWC data \cite{Abeysekara:2017old}. The diffusion coefficient \texttt{D0} and energy conversion efficiency \texttt{eta} are determined by fitting the data (see figure caption), while the other parameters are set to default values as listed in Table~\ref{tab:params}.
Ref.~\cite{Wang:2021xph,Fang:2021qon} have noted that the superdiffusion model with $\alpha\gtrsim1.5$ can provide good agreement with the current measurements, yet it has fundamentally different characteristics compared to standard diffusion. As illustrated in Fig.~\ref{fig:prof}, superdiffusion results in a more contracted shape at small angles and a more extended shape at large angles. Due to this distinct feature, it is feasible to investigate the superdiffusion phenomenon as future experiments become more precise.

\begin{figure}[!htb]
\centering
\includegraphics[width=0.48\textwidth]{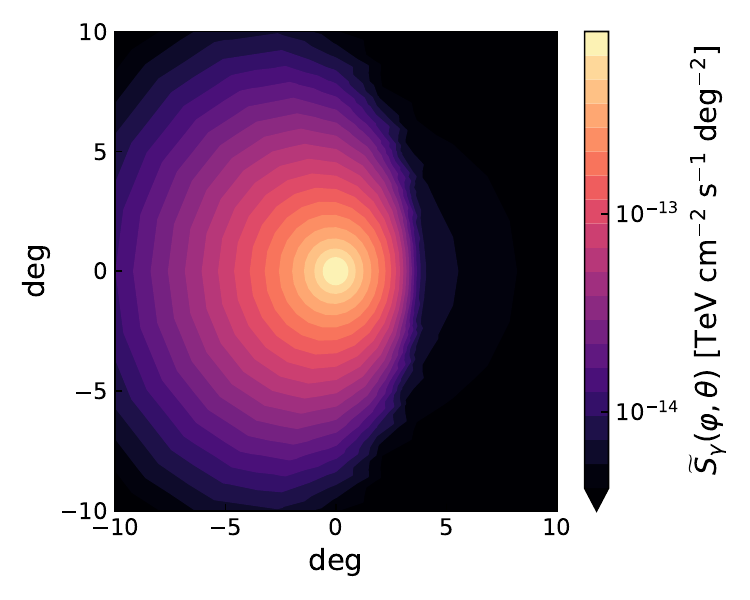}
\includegraphics[width=0.48\textwidth]{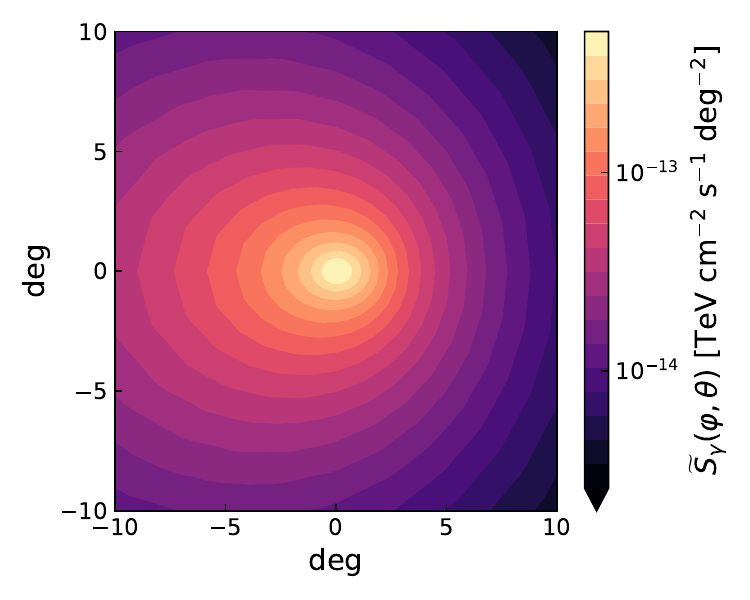}
\caption{Two-dimensional $\gamma$SB of the Geminga halo predicted by models \texttt{SNR2Z\_N} (left) and \texttt{Aniso\_N} (right). The parameters used are the default settings in \texttt{param\_config.yaml}, with the following exceptions: \texttt{PHI=270} for the \texttt{SNR2Z\_N} model (indicating the pulsar motion is horizontally to the right); \texttt{z\_ref=300} for the \texttt{Aniso\_N} model.}
\label{fig:map}
\end{figure}

The files \texttt{[root]\_SBmap\_bin*.dat} represent the 2D $\gamma$SB of cylindrically symmetric models after PSF convolution. Figure~\ref{fig:map} illustrates the $\widetilde{S}_\gamma(\varphi,\theta)$ of the Geminga halo as derived from the \texttt{SNR2Z\_N} model and the \texttt{Aniso\_N} model, respectively, both of which predict asymmetries in the Geminga halo. For the \texttt{SNR2Z\_N} model, the halo asymmetry arises from the significant eccentricity of the pulsar within the SNR. For the \texttt{Aniso\_N} model, the asymmetry is due to the slight misalignment between the mean field direction around the pulsar and the LOS of the observer. If the correlation length of the ISM magnetic field is taken into account, the asymmetry predicted by the \texttt{Aniso\_N} model could be more complex \cite{Fang:2023axu,2024arXiv240702478B,2025ApJ...987...19Y}. Currently, the HAWC observation of the Geminga halo has preliminarily indicated the presence of morphological asymmetry; the difference in effective diffusion coefficients among sectors has exceeded a significance of $3\sigma$ \cite{HAWC:2024scl}.

\begin{figure}[!htb]
\centering
\includegraphics[width=0.48\textwidth]{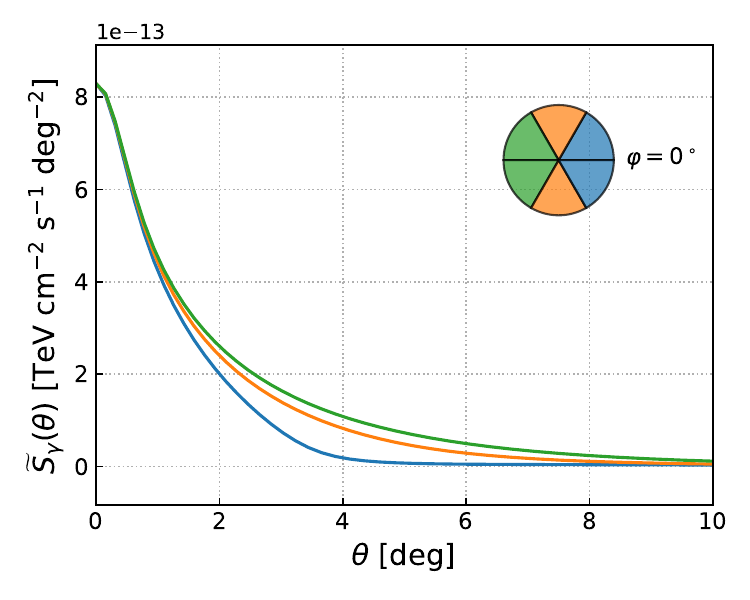}
\includegraphics[width=0.48\textwidth]{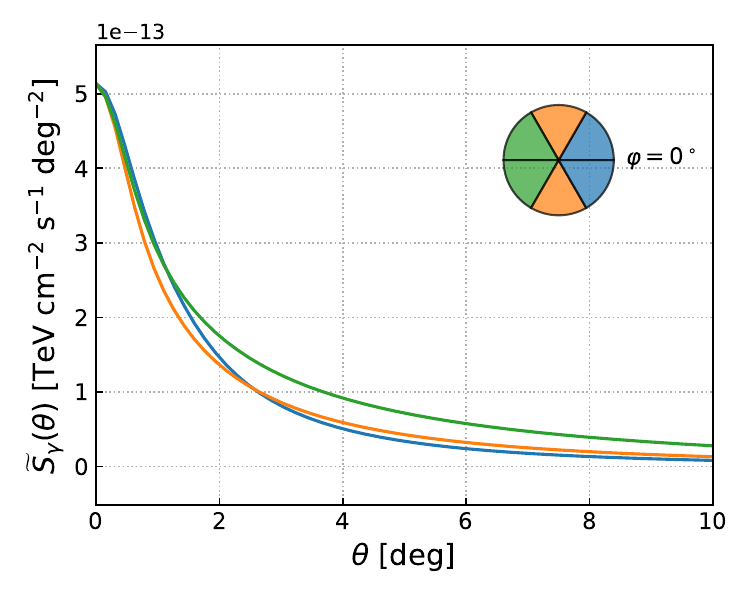}
\caption{Average of $\gamma$SB over different azimuthal intervals for models \texttt{SNR2Z\_N} (left) and \texttt{Aniso\_N} (right). The parameters used are identical to those in Fig.~\ref{fig:map}.}
\label{fig:prof2d}
\end{figure}

Unlike spherically symmetric models, the files \texttt{[root]\_SBprof\_bin*.dat} generated by cylindrically symmetric models contain the average of $\widetilde{S}_\gamma(\theta)$ over different azimuthal intervals. The number of intervals is determined by \texttt{NUM\_sect}, which divides the range $\varphi=0-360^\circ$ into \texttt{NUM\_sect} sectors. Figure~\ref{fig:prof2d} illustrates the average 1D $\gamma$SB across different azimuthal intervals for the two models corresponding to Fig.~\ref{fig:map}. The models exhibit unique characteristics: the \texttt{SNR2Z\_N} model shows a monotonic variation in extension for $\varphi=0-180^\circ$, whereas the \texttt{Aniso\_N} model displays a more complex pattern of morphological changes. These differences can be tested through more precise measurements in the future.

The file \texttt{[root]\_spec.dat} contains the $\gamma$-ray energy spectrum of the pulsar halo within an angle $\theta_b$ around the pulsar, i.e., $F(E_\gamma)= \int_0^{\theta_b} s_\gamma(E_\gamma,\theta)2\pi\theta d\theta$, where $\theta_b$ is determined by \texttt{tht\_bound} in the configuration file. In practice, $\gamma$-ray experiments perform reliable measurements of the energy spectrum within specific analysis regions. We can select the $\theta_b$ corresponding to the designated analysis region to perform model-data comparisons.

\begin{figure}[!htb]
\centering
\includegraphics[width=0.6\textwidth]{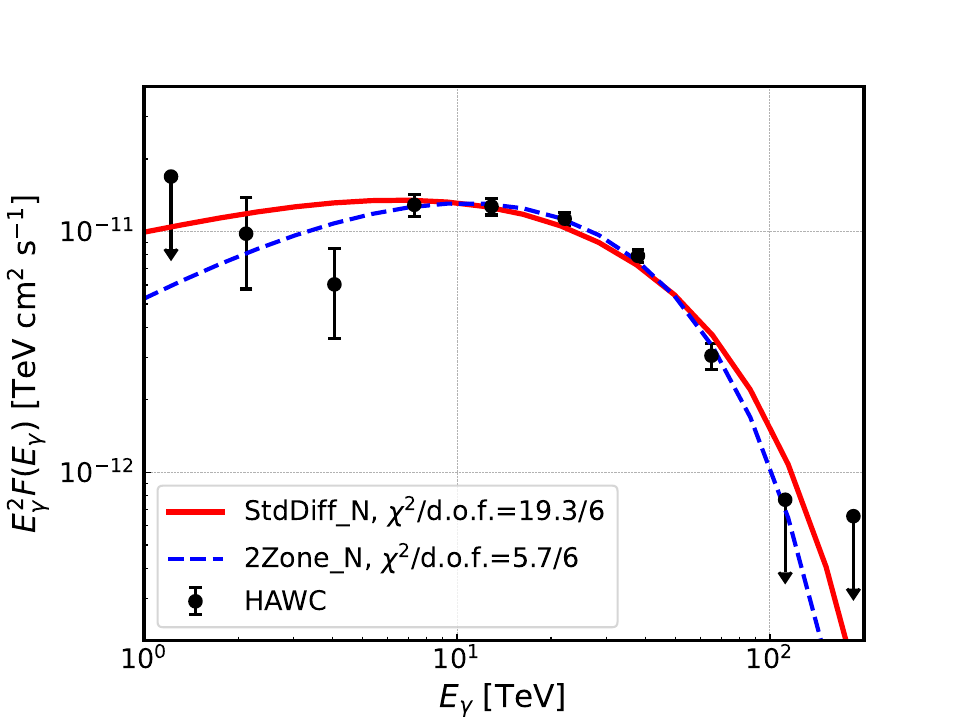}
\caption{$\gamma$-ray spectra of the Geminga halo within $10^\circ$ around the pulsar computed using the \texttt{StdDiff\_N} and \texttt{2Zone\_N} models, in comparison with the HAWC data. For the \texttt{StdDiff\_N} model: $\texttt{eta=0.06}$, $\texttt{Ec=165}$, and $\texttt{D0=4.5e27}$. For the \texttt{2Zone\_N} model: $\texttt{eta=0.09}$, $\texttt{Ec=125}$, $\texttt{D0=7e27}$, and $\texttt{r\_2z=35}$, where the last two parameters are suggested by Ref.~\cite{Fang:2023xla}. Other parameters are set to their default values.}
\label{fig:spec}
\end{figure}

In Fig.~\ref{fig:spec}, we present the $\gamma$-ray spectra of the Geminga halo with $\theta_b=10^\circ$ computed using the \texttt{StdDiff\_N} and the \texttt{2Zone\_N} models, respectively. The spectrum measured by HAWC \cite{HAWC:2024scl} is shown for comparison\footnote{The energy spectrum provided by HAWC is extrapolated to $\theta_b=30^\circ$ assuming the standard diffusion model \cite{HAWC:2024scl}. However, since the actual analysis region is only $\sim10^\circ$ around the pulsar, the flux in the extrapolated region is model-dependent and therefore not reliable. As a result, the data we are using is rescaled from the energy spectrum provided by HAWC to $\theta_b=10^\circ$ for comparison with the model. We also suggest that experimental groups provide direct spectrum measurements within the analysis region, rather than extrapolated results based on models.}. The pronounced curvature observed in the spectrum could be interpreted by a hard electron injection spectrum, while acceleration theories of bow-shock PWNe suggest that \texttt{index\_p} should not be significantly smaller than $1.0$ \cite{Bykov:2017xpo}. Therefore, we set $\texttt{index\_p}=1.0$ for both models and determine the remaining injection parameters (\texttt{eta} and \texttt{Ec}) by fitting the HAWC data. The upper limits at the highest and lowest energy bins have very weak constraints on the model and are not included in the fitting process.

It can be seen that under the same setting of \texttt{index\_p}, the \texttt{2Zone\_N} model can better reproduce the observed spectral curvature, thereby achieving significantly improved goodness of fitting compared to the \texttt{StdDiff\_N} model. Under a two-zone diffusion scenario, because lower energy electrons have a larger characteristic diffusion distance, a greater proportion of them escape from the slow-diffusion zone. Consequently, there is a reduced fraction of $\gamma$-ray flux within $10^\circ$ at lower energies, which leads to a harder spectrum at low energies. One may refer to Ref.~\cite{Fang:2023xla} for a detailed discussion.

In addition to the default output files mentioned above, we provide the energy-differential $\gamma$SB prior to the PSF convolution, namely $s_\gamma(E_\gamma,\theta)$ or $s_\gamma(E_\gamma,\varphi,\theta)$, in the form of a structure variable (named \texttt{SB\_out}) within the interface program \texttt{Phect.c}. This allows users to conduct more sophisticated PSF convolutions beyond the Gaussian model. Users can also perform custom calculations based on $s_\gamma$ or incorporate it into model-fitting programs.

\section{Verification tests}
\label{sec:test}
We assess the reliability of the computation results through comparisons from both internal and external perspectives.

\texttt{StdDiff\_A} is among the simplest models in \texttt{PHECT}, and its solution via a semi‑analytical approach is relatively straightforward. We therefore adopt it as a benchmark to test whether more complex models can reproduce results consistent with \texttt{StdDiff\_A} under appropriate limiting parameters. This procedure serves to verify the fundamental correctness of the complex models, particularly those based on numerical schemes.

We adopt the default parameters from \texttt{param\_config.yaml} for the \texttt{StdDiff\_A} model. The \texttt{StdDiff\_N} and \texttt{StdDiff\_A0} models share the same physics as \texttt{StdDiff\_A} and also use the default parameters. The superluminal correction in the \texttt{Juttner\_A} model becomes significant when \texttt{D0} is large, but should reduce to \texttt{StdDiff\_A} under default parameters. The diffusion coefficient for the \texttt{Super\_A} model has the units of $\text{cm}^{\alpha}\ \text{s}^{-1}$, which carries a different physical meaning from normal diffusion. However, as $\alpha$ approaches $2$ (e.g., $\alpha=1.9999$), the expected results of \texttt{Super\_A} should be consistent with standard diffusion. The anisotropic diffusion models \texttt{Aniso\_A} and \texttt{Aniso\_N} reduce to standard diffusion when $\texttt{Ma}=1$. The spatially dependent diffusion model \texttt{SNR2Z\_N} reduces to standard diffusion when $\texttt{ratio\_snr}=1$.

\begin{figure}[!htb]
\centering
\includegraphics[width=0.56\textwidth]{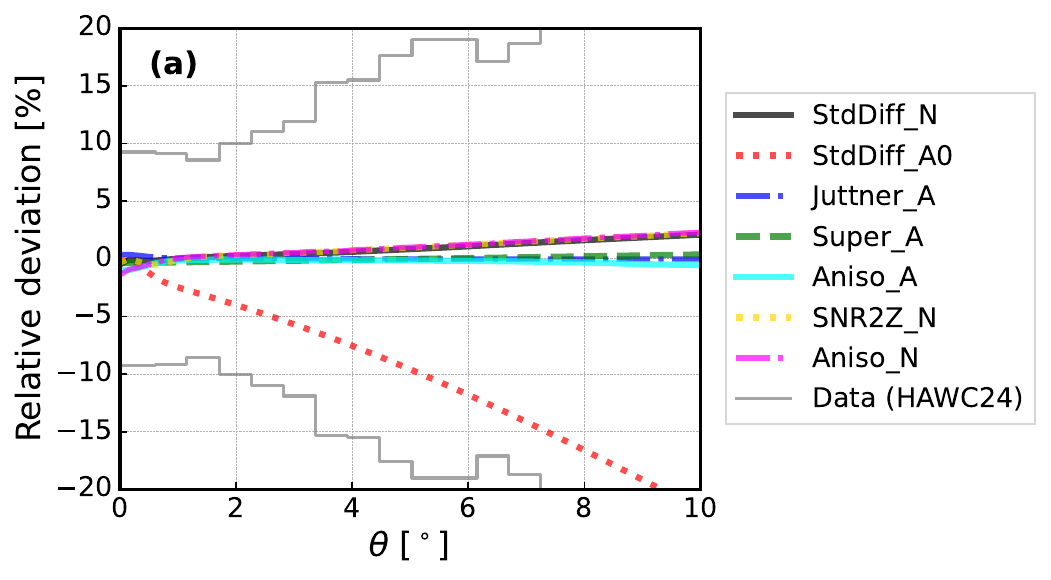}
\includegraphics[width=0.4\textwidth]{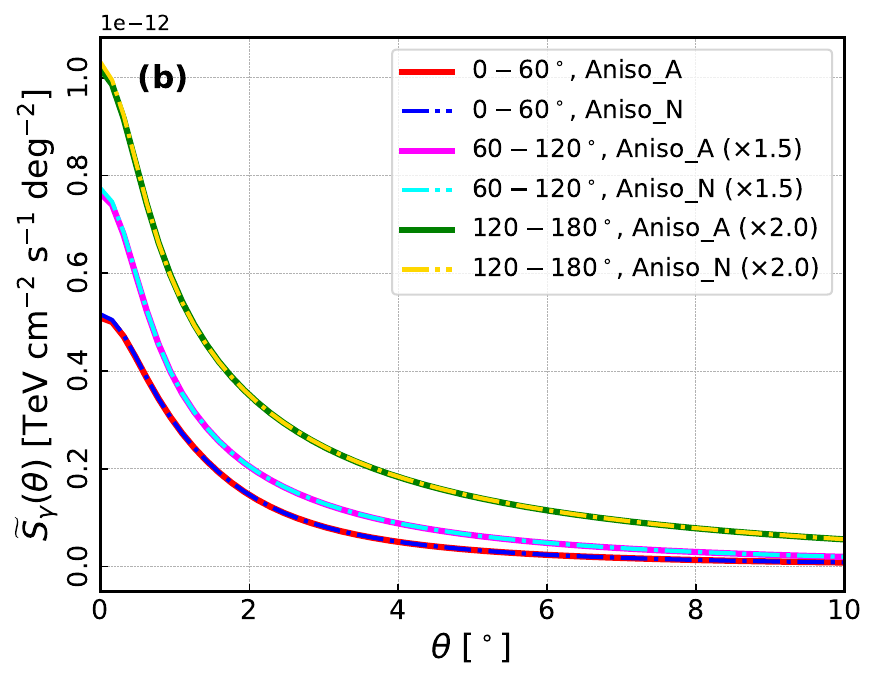}
\caption{Internal verification tests of \texttt{PHECT}. (a) Relative deviation of the 1D $\gamma$SB in $8-40$~TeV for different models with respect to \texttt{StdDiff\_A}. Each model can reduce to \texttt{StdDiff\_A} under appropriate limiting parameters; this consistency serves as a fundamental reliability test for the complex models. See the main text for detailed parameter settings. The relative measurement uncertainty of the Geminga halo from HAWC \cite{HAWC:2024scl} is shown for comparison. (b) Comparison of $\gamma$SB across different azimuthal intervals computed by \texttt{Aniso\_A} and \texttt{Aniso\_N}, with parameters set to $\texttt{Ma}=0.2$ and $\texttt{PHI}=3$.}
\label{fig:consistency}
\end{figure}

Figure~\ref{fig:consistency}(a) presents the relative deviation of the 1D $\gamma$SB from different models with respect to \texttt{StdDiff\_A}, i.e., $(\widetilde{S}_\gamma-\widetilde{S}_{\gamma,0})/\widetilde{S}_{\gamma,0}$, where $\widetilde{S}_\gamma$ and $\widetilde{S}_{\gamma,0}$ are the results of the specified model and the \texttt{StdDiff\_A} model, respectively\footnote{The \texttt{2Zone\_N} model is not shown as it is identical to \texttt{StdDiff\_N} for $\texttt{ratio\_2z=1.0}$.}. Since the default parameter setting is referred to Geminga, we also display the measurement relative error of HAWC on the Geminga halo for comparison \cite{HAWC:2024scl}. It can be seen that, except for \texttt{StdDiff\_A0}, the relative deviations of all models are within $3\%$ and significantly smaller than the experimental error. \texttt{StdDiff\_A0} exhibits a systematic bias as it underestimates the characteristic diffusion distance of electrons (see Eq.~\ref{eq:lambda3}).

For modeling anisotropic diffusion, \texttt{PHECT} provides both a semi-analytical and a numerical model. While the computational speed of \texttt{Aniso\_A} is significantly slower than that of \texttt{Aniso\_N}, it offers a more straightforward solution procedure. We compare the results from these models to verify the robustness of \texttt{Aniso\_N} under conditions where $\texttt{Ma}\neq1$. Figure \ref{fig:consistency}(b) shows the $\gamma$SB over different azimuthal intervals computed by the two models, with parameters set to $\texttt{Ma}=0.2$, $\texttt{PHI}=3$, and $\texttt{z\_ref}=300$\footnote{For the \texttt{Aniso\_N} model, it is advisable to set a larger \texttt{z\_ref} value, which correspondingly increases the basic step size in the $z$ direction.}. As shown, the numerical solution exhibits excellent agreement with the semi-analytical solution.

\begin{figure}[!htb]
\centering
\includegraphics[width=0.45\textwidth]{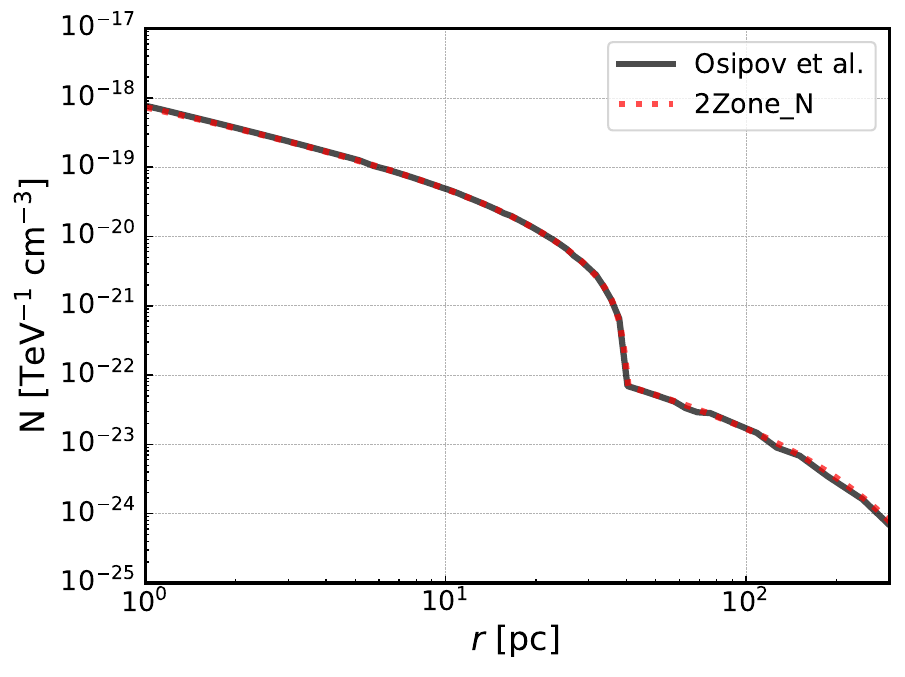}
\caption{Comparison of the electron number density distribution (at $\approx50$~TeV) computed by the \texttt{2Zone\_N} model and the semi-analytical model proposed by \citet{Osipov:2020lty}. The parameter settings are kept at default values (including \texttt{r\_2z=40} and \texttt{ratio\_2z=100}), except for $\texttt{NUM\_r}=100$ and $\texttt{r\_ref}=70$.}
\label{fig:2zone}
\end{figure}

For the two-zone diffusion problem within the framework of pulsar halos, in addition to the numerical approach using the finite volume method employed by the \texttt{2Zone\_N} model, alternative methods include a numerical technique based on stochastic diffusion equations as discussed in Ref.~\cite{Profumo:2018fmz}, and semi-analytical solutions proposed in Ref.~\cite{Tang:2018wyr,Osipov:2020lty}. However, Ref.~\cite{Osipov:2020lty} points out that Ref.~\cite{Tang:2018wyr} incorrectly treats the continuity condition at $r=\texttt{r\_2z}$, leading to notable deviations in the results. Therefore, to verify the accuracy of our \texttt{2Zone\_N} model, we compare its predicted electron number density distribution with the semi-analytical solution from Ref.~\cite{Osipov:2020lty}. As illustrated in Fig.~\ref{fig:2zone}, the two show excellent agreement.

In the numerical models within \texttt{PHECT}, the \texttt{SNR2Z\_N} model is currently the only one without a direct semi-analytical counterpart for verification. Nevertheless, its computational framework follows the same logic as the validated \texttt{2Zone\_N} model, extending from spherical to cylindrical symmetry. Given the established accuracy of the \texttt{2Zone\_N} model, the reliability of the \texttt{SNR2Z\_N} model is also substantiated to a considerable extent.

\begin{figure}[!htb]
\centering
\includegraphics[width=0.5\textwidth]{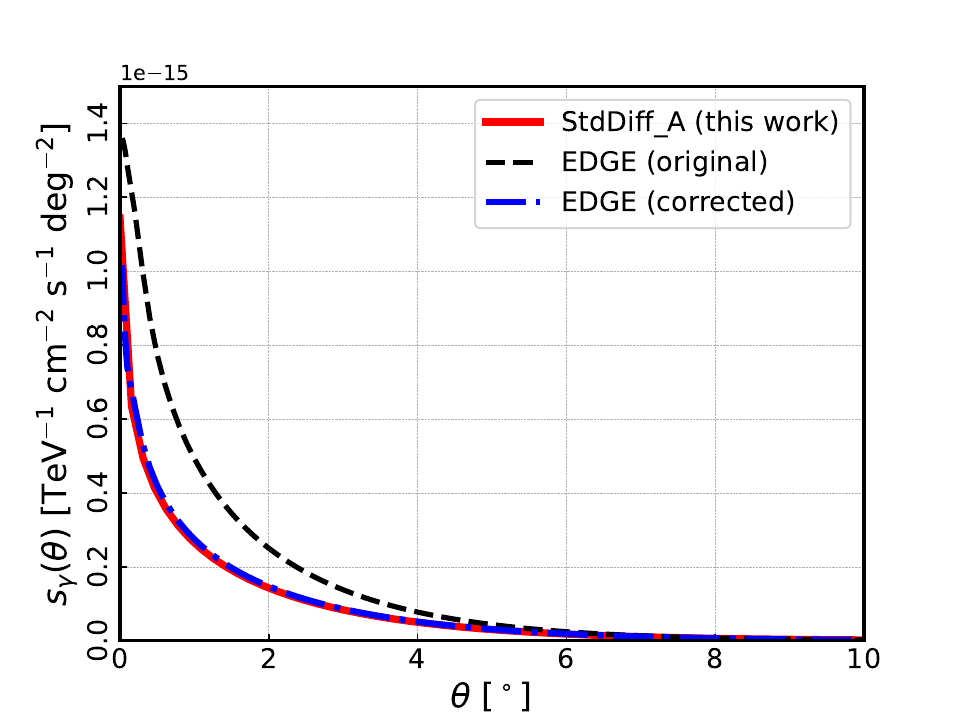}
\caption{Comparison of the energy differential 1D $\gamma$SB (at $20$~TeV) of the Geminga halo computed by the \texttt{StdDiff\_A} model and the external computational tool \texttt{EDGE} \cite{2017ICRC...35..735L}. After correcting several implementation errors in the \texttt{EDGE} code, the two exhibit good consistency. See Appdenix~\ref{app:edge} for a detailed explanation of the corrections.}
\label{fig:edge}
\end{figure}

Finally, we compare our benchmark model \texttt{StdDiff\_A} with the results from the external computational tool \texttt{EDGE} \cite{2017ICRC...35..735L}. \texttt{EDGE} can compute the ICS $\gamma$-ray emission from pulsar halos under standard diffusion, with good transparency on the computation process and parameter settings, making it suitable for cross-validation. The comparison results are presented in Fig.~\ref{fig:edge}, which is the energy differential $\gamma$SB of the Geminga halo with parameters set to the default values of \texttt{EDGE}.

During the validation exercise, we identified several implementation errors in the \texttt{EDGE} code. The corresponding discussion and corrections are detailed in Appendix~\ref{app:edge}. The major issue is that the \texttt{EDGE} code does not reset the background photon field at each initialization, leading to an erroneous repeated summation of background photon energy density and consequently an overestimation of the $\gamma$-ray flux (as shown by the black dashed line in Fig.~\ref{fig:edge}). After applying the necessary corrections, \texttt{EDGE} produces a $\gamma$SB profile consistent with that computed by the \texttt{StdDiff\_A} model.

\section{Conclusion and perspective}
\label{sec:conclu}
This work introduces \texttt{PHECT}, a lightweight software designed for pulsar halo emission modeling. Compared to existing tools (e.g., \texttt{EDGE} and \texttt{GALPROP}), the characteristics of \texttt{PHECT} are summarized as follows: 
\begin{itemize} 
 \item User-friendly design: This tool is highly targeted and requires minimal dependencies, allowing users to perform necessary computations simply by configuring an intuitive \texttt{YAML} file.
 \item Diverse propagation models: The tool incorporates multiple theoretical and phenomenological transport models beyond homogeneous and isotropic diffusion.
 \item More suitable numerical methods: For models involving numerical computation, the tool employs the finite volume method to discretize the propagation equation, which remains robust when encountering discontinuous coefficients and non-uniform grids. 
 \item Low update cost: Given its compact size, the tool enables significantly shorter update intervals.
 \end{itemize} 
With the data accumulation from experiments such as LHAASO, HAWC, and H.E.S.S., both new pulsar halos and finer features of known halos (e.g., the Geminga halo) are emerging. Compared to the parameterized formulas commonly used to describe pulsar halo emission, \texttt{PHECT} facilitates the transformation of precise measurements into physically meaningful constraints and enables direct model comparisons through self-consistent computations.

In addition to incorporating the self-excited model and improving the superdiffusion model as mentioned in Sec.~\ref{sec:model}, \texttt{PHECT} also aims to introduce further features. The current version does not consider pulsar motion, as its impact on TeV-band halo morphology is expected to be negligible \cite{DiMauro:2019yvh,ZhangZhangYi:2021kzq}. In the GeV energy range, however, pulsar motion may induce a significant shift of the halo center relative to the pulsar \cite{Johannesson:2019jlk}. In upcoming updates, we plan to incorporate models that account for pulsar motion. Currently, the Fermi collaboration has reported several potential GeV pulsar halos in their search for extended sources \cite{2024arXiv241107162A}, none of which overlap with the known TeV pulsar halos. Among the detected TeV halos, only the Geminga halo has been reported to have a potential GeV counterpart \cite{DiMauro:2019yvh}, although the detection remains debated \cite{Shao-Qiang:2018zla}.

The high-energy electrons that generate $\gamma$-ray halos through ICS also produce x rays through synchrotron radiation in the magnetic field. However, current x-ray experiments, including the wide-field eROSITA, have yet to detect x-ray emission corresponding to pulsar halos\footnote{Recent studies using eROSITA have detected extended x-ray signals within $0.2^\circ$ around Monogem \cite{2025arXiv250117046N}, potentially indicating localized magnetic field enhancement. However, alternative interpretations suggest these signals could originate from leakage of the Monogem pulsar itself \cite{Khokhriakova:2025rhn}.} \cite{Liu:2019sfl,Khokhriakova:2023rqh,Manconi:2024wlq,2025arXiv251213846K}. This might be due to a relatively weak root-mean-square magnetic field in the ISM within the halo regions\footnote{Even if the root-mean-square magnetic field could be weak, the system can still be in a state of strong turbulence as long as the turbulent magnetic field dominates the magnetic energy.}, or because the mean field near the pulsar, as suggested by the anisotropic diffusion model, is oriented close to the LOS. In future versions, we plan to incorporate computations of synchrotron emission.

The standard treatment of electron energy loss through ICS adopts the continuous approximation $|dE_e/dt|=b_{0,\mathrm{ics}}E_e^2$, which describes the mean energy evolution of electron populations. However, ICS is fundamentally a discrete process involving random scattering events with randomized energy losses per interaction, resulting in non-monochromatic electron energy distributions rather than single-valued trajectories \cite{John:2022asa,Xia:2025dow}. Consequently, the discrete ICS correction predicts broader $\gamma$-ray spectral structures than the continuous approximation, with increasing discrepancies at higher energies \cite{Xia:2025dow}. Future versions of \texttt{PHECT} will incorporate exact ICS energy loss computations.

\begin{acknowledgments}
The author thanks Dr.~Yue Zhu for testing and correcting the EDGE code. This work is supported by the National Natural Science Foundation of China under Grants No. 12393853, No. 12105292, and the National Key R\&D program of China under Grant No. 2024YFA1611401.
\end{acknowledgments}

\appendix

\section{Adjustable parameters}
\label{app:params}

\begin{ThreePartTable}
\begin{TableNotes}
  \item[a] The units are different for the \texttt{Super\_A} model.
  \item[b] For the \texttt{Aniso\_N} model, $D_{zz}$ could be quite large, thus \texttt{z\_ref} needs to be increased accordingly.
  \item[c] The actual number of grid points in the $z$ direction is $2\mathtt{NUM\_z}-1$. For more details, please refer to Appendix~\ref{app:cylin}.
  \item[d] See text.
  \item[e] The size of these arrays is \texttt{NUM\_energy}.
\end{TableNotes}

\begin{longtable}{lcc}
  \caption{Adjustable parameters (in \texttt{param\_config.yaml})} \label{tab:params} \\
  \toprule
  Name & Description & Default \\ \midrule
  \endfirsthead
  \multicolumn{3}{c}{Table~\ref{tab:params} (Continued)} \\ \toprule
  Name & Description & Default \\ \midrule  
  \endhead
  \bottomrule
  \endfoot
  \insertTableNotes
  \endlastfoot
    
  \texttt{ts\_c} & Pulsar characteristic age [yr] & \texttt{3.42e5} \\
  \texttt{rs} & Pulsar distance [pc] & \texttt{250} \\ 
  \texttt{Edot\_now} & Current spin-down luminosity of pulsar [erg] & \texttt{3.2e34} \\
  \texttt{tau0} & Initial spin-down timescale [yr] & \texttt{1e4} \\
  \midrule
  \texttt{eta} & Conversion efficiency from spin-down energy to $e^\pm$ energy & \texttt{0.1} \\
  \texttt{index\_p} & Injection spectrum: Power-law index & \texttt{1.0} \\
  \texttt{Ec} & Injection spectrum: Cutoff energy [TeV] & \texttt{100} \\
  \texttt{index\_exp} & Injection spectrum: Super-exponential index & \texttt{2.0} \\ 
  \midrule
  \texttt{model} & Name of electron propagation model & \texttt{StdDiff\_N} \\
  \texttt{D0} & Diffusion coefficient at $100$~TeV [cm$^2$~s$^{-1}$]\tnote{a} & \texttt{1e28} \\
  \texttt{delta} & Energy index of diffusion coefficient & \texttt{0.333} \\
  \texttt{alpha} & L\'{e}vy flight index (for \texttt{Super\_A}) & \texttt{2.0} \\
  \texttt{r\_2z} & Slow-diffusion zone size [pc] (for \texttt{2Zone\_N}) & \texttt{40} \\
  \multirow{2}{*}{\texttt{ratio\_2z}} & Ratio of $D_0$ outside to inside slow-diffusion zone & \multirow{2}{*}{\texttt{100}} \\
  & (for \texttt{2Zone\_N}) & \\
  \multirow{2}{*}{\texttt{PHI}} & Angle between z-axis and the vector to observer [deg] & \multirow{2}{*}{\texttt{3}} \\
  & (for cylindrical symmetric models) & \\
  \multirow{2}{*}{\texttt{Ma}} & Alfv\'{e}nic Mach number of turbulent magnetic field & \multirow{2}{*}{\texttt{0.2}}  \\
  & (for \texttt{Aniso\_N} and \texttt{Aniso\_A}) & \\
  \multirow{2}{*}{\texttt{z\_snr}} & z coordinate of SNR center [pc] & \multirow{2}{*}{\texttt{-75}} \\
  & (for \texttt{SNR2Z\_N}) & \\
  \texttt{R\_snr} & Current SNR size [pc] (for \texttt{SNR2Z\_N}) & \texttt{90} \\
  \multirow{2}{*}{\texttt{ratio\_snr}} & Ratio of $D_0$ outside to inside SNR & \multirow{2}{*}{\texttt{100}} \\
  & (for \texttt{SNR2Z\_N}) & \\
  \midrule
  \texttt{dt0} & Basic time step for numerical computation [yr] & \texttt{50} \\
  \texttt{Ee\_min} & Min. energy of computed electron number density [TeV] & \texttt{1.0} \\
  \texttt{Ee\_max} & Max. energy of computed electron number density [TeV] & \texttt{1000} \\
  \texttt{NUM\_Ee} & Number of grid points of $E_e$ & \texttt{50} \\
  \texttt{r\_max} & Max. $r$ to pulsar [pc] & \texttt{3000} \\
  \texttt{r\_ref} & Where $r$ step amplifies \texttt{r\_ampl} times [pc] & \texttt{30} \\
  \texttt{r\_ampl} & Amplification of $r$ step at \texttt{r\_ref} & \texttt{3.0} \\
  \texttt{NUM\_r} & Number of grid points of $z$ & \texttt{60} \\
  \texttt{z\_max} & Max. $z$ to pulsar [pc] & \texttt{3000} \\
  \texttt{z\_ref\tnote{b}} & Where $z$ step amplifies \texttt{z\_ampl} times [pc] & \texttt{30} \\
  \texttt{z\_ampl} & Amplification of $z$ step at \texttt{z\_ref} & \texttt{3.0} \\
  \texttt{NUM\_z\tnote{c}} & Number of grid points of $z$ & \texttt{60} \\
  \midrule
  \texttt{Eg\_min} & Min. energy of computed $\gamma$SB [TeV] & \texttt{2.0} \\
  \texttt{Eg\_max} & Max. energy of computed $\gamma$SB [TeV] & \texttt{200} \\
  \texttt{NUM\_Eg} & Number of grid points of $E_\gamma$ & \texttt{15} \\
  \texttt{tht\_max} & Max. angular distance to pulsar of computed $\gamma$SB [pc]\tnote{d} & \texttt{100} \\
  \texttt{tht\_ref} & Where $\theta$ step amplifies \texttt{tht\_ampl} times [pc]\tnote{d} & \texttt{50} \\
  \texttt{tht\_ampl} & Amplification of $\theta$ step at \texttt{tht\_ref} & \texttt{5.0} \\
  \texttt{NUM\_tht} & Number of grid points of $\theta$ & \texttt{50} \\
  \texttt{tht\_bound} & Upper limit of the integral in Eq.~(\ref{eq:spec}) [pc]\tnote{d} & \texttt{50} \\
  \texttt{los\_bound} & Max. distance from pulsar in LOS integration [pc] & \texttt{200} \\
  \texttt{NUM\_phi} & Determine $\varphi$ step in computations: $\Delta\varphi=180^\circ/\mathtt{NUM\_phi}$ & \texttt{12} \\
  \texttt{NUM\_sect} & Number of output azimuth sectors ($0-360^\circ$) & \texttt{6} \\
  \texttt{NUM\_energy} & Number of output energy bins of $\gamma$SB & \texttt{3} \\
  \texttt{sgm\_psf} & Width of the 2D Gaussian PSF [deg] & \texttt{[0.3, 0.2, 0.15]}\tnote{e} \\
  \texttt{Eg\_1} & Lower bounds of output energy bins of $\gamma$SB [TeV] & \texttt{[8, 40, 100]}\tnote{e} \\
  \texttt{Eg\_2} & Upper bounds of output energy bins of $\gamma$SB [TeV] & \texttt{[40, 100, 150]}\tnote{e} \\
  \midrule
  \texttt{B} & Magnetic field strength in ISM [$\mu$G] & \texttt{3.0} \\
  \texttt{T\_cmb} & Temperature of CMB [K] & \texttt{2.275} \\
  \texttt{T\_dust} & Temperature of dust emission [K] & \texttt{20} \\
  \texttt{T\_sl} & Temperature of starlight [K] & \texttt{5000} \\
  \texttt{density\_cmb} & Energy density of CMB [eV~cm$^{-3}$] & \texttt{0.26} \\
  \texttt{density\_dust} & Energy density of dust emission [eV~cm$^{-3}$] & \texttt{0.3} \\
  \texttt{density\_sl} & Energy density of starlight [eV~cm$^{-3}$] & \texttt{0.3} \\
  \midrule
  \texttt{switch\_gamma} & Switch for $\gamma$-ray emission computation & \texttt{true} \\
  \texttt{gunit} & Energy unit of $S_\gamma$ & \texttt{"energy"} \\
  \texttt{root} & Root name of output files & \texttt{"default"} \\
  \bottomrule
\end{longtable}
\end{ThreePartTable}

By definition, \texttt{tht\_max}, \texttt{tht\_ref}, and \texttt{tht\_bound} are specified in degrees. However, given the varying distances to pulsars, physical scales are more practical for configuring these parameters. For example, the value of \texttt{tht\_max} is specified in parsecs in the configuration file \texttt{param\_config.yaml}. If we denote this value as $x$, the actual \texttt{tht\_max} used in computations is then $x/\mathtt{rs}\cdot180^\circ/\pi$.

\section{Semi-analytical solutions to electron propagation}
\label{app:ana}
Certain propagation models in \texttt{PHECT} use semi-analytical solutions based on the Green's function method. This section presents the explicit forms of these solutions without detailed derivations. Numerical integrations involved in the semi-analytical solutions are performed using the \texttt{CQUAD} algorithm in \texttt{GSL}.

\subsection{\texttt{StdDiff\_A} and \texttt{StdDiff\_A0} models}
\label{app:norm}
The general solution to Eq.~(\ref{eq:prop}) using the Green's function method is given by
\begin{equation}
 N(E_e, \mathbi{r}, t)=\int_{R^3}d^3\mathbi{r}_0\int dE_{e,0}\int dt_0\,Q(E_{e,0},\mathbi{r}_0,t_0)\,G(E_e,\mathbi{r},t;E_{e,0},\mathbi{r}_0,t_0)\,.
 \label{eq:green}
\end{equation}
The \texttt{StdDiff\_A} and \texttt{StdDiff\_A0} models are based on standard (homogeneous and isotropic) diffusion, corresponding to the Green's function (e.g., see Ref.~\cite{1964ocr..book.....G}):
\begin{equation}
 G(E_e,\mathbi{r},t;E_{e,0},\mathbi{r}_0,t_0)=\frac{1}{b(E_e)(4\pi\lambda)^{3/2}}\,\exp\left[-\frac{(\mathbi{r}-\mathbi{r}_0)^2}{4\lambda}\right]\,\delta(t-t_0-\tau)H(\tau)\,,
 \label{eq:green_norm}
\end{equation}
where $H$ is the step function, and
\begin{equation}
 \tau=\int_{E_e}^{E_{e,0}}\frac{dE_e^\prime}{b(E_e^\prime)}\,,
 \label{eq:tau}
\end{equation}
\begin{equation}
 \lambda=\int_{E_e}^{E_{e,0}}\frac{D(E_e^\prime)}{b(E_e^\prime)}dE_e^\prime\,.
 \label{eq:lambda}
\end{equation}
By combining the form of the source function introduced in Sec.~\ref{subsec:electron}, we can derive from Eq.~(\ref{eq:green}) that
\begin{equation}
 N(E_e, r, t_s)=\int_{t_\mathrm{min}}^{t_s}dt_0\,q_t(t_0)\,q_E(E_{e,\star})\,\frac{b(E_{e,\star})}{b(E_e)}\,\frac{1}{(4\pi\lambda_\star)^{3/2}}\,\exp\left(-\frac{r^2}{4\lambda_\star}\right)\,,
 \label{eq:sol_norm}
\end{equation}
where 
\begin{equation}
 \lambda_\star=\int_{E_e}^{E_{e,\star}}\frac{D(E_e^\prime)}{b(E_e^\prime)}dE_e^\prime\,,
 \label{eq:lambda2}
\end{equation}
and $E_{e,\star}$ is the root of
\begin{equation}
 t_s-t_0-\tau=0\,.
 \label{eq:root}
\end{equation}
By definition,
\begin{equation}
 t_\mathrm{min}=t_\mathrm{cool}\equiv t_s-\int_{E_e}^{\infty}\frac{dE_e^\prime}{b(E_e^\prime)}\,.
 \label{eq:tmin}
\end{equation}
However, since the source function vanishes for $t_0<0$, we set $t_\mathrm{min}=\mathrm{max}\{t_\mathrm{cool}, 0\}$ in practice to avoid an extended null region in the integrand.

The \texttt{StdDiff\_A} model performs accurate computation for Eq.~(\ref{eq:sol_norm}), requiring the solution of an integral equation, i.e. Eq.~(\ref{eq:root}), to obtain $E_{e,\star}$, along with a numerical integration for Eq.~(\ref{eq:lambda2}). As a result, the \texttt{StdDiff\_A} model demands longer computational time. For the solution of Eq.~(\ref{eq:root}), we adopt Brent's root-finding method embedded in \texttt{GSL}. The models \texttt{Juttner\_A} and \texttt{Super\_A} obtain $E_{e,\star}$ and $\lambda_\star$ using identical methods.

The \texttt{StdDiff\_A0} model employs approximations for the above computation \cite{Delahaye:2010ji,Kobayashi:2003kp}, which can significantly enhance efficiency when a rough estimate is sufficient. Equation~(\ref{eq:tau}) is approximated by
\begin{equation}
 \tau\approx\frac{1}{b_0(E_e)}\int_{E_e}^{E_{e,0}}\frac{dE_e^\prime}{{E_e^\prime}^2}=\frac{1}{b_0(E_e)}\left(\frac{1}{E_e}-\frac{1}{E_{e,0}}\right)\,,
 \label{eq:tau2}
\end{equation}
and the root of Eq.~(\ref{eq:root}) is
\begin{equation}
 E_{e,\star}\approx\frac{E_e}{1-b_0(E_e)E_e(t_s-t_0)}\,.
 \label{eq:EE}
\end{equation}
Similarly, we have
\begin{equation}
 \lambda_\star\approx\frac{1}{b_0(E_e)}\int_{E_e}^{E_{e,\star}}\frac{D(E_e^\prime)}{{E_e^\prime}^2}dE_e^\prime\,,
 \label{eq:lambda3}
\end{equation}
where numerical integration is no longer necessary when $D(E_e)$ is in a power-law form \cite{Kobayashi:2003kp}. 

Due to the Klein-Nishina effect, Eq.~(\ref{eq:tau2}) and (\ref{eq:lambda3}) overestimate the energy-loss rate, which results in underestimating both the final $\gamma$-ray flux and the extension of the source. This discrepancy becomes more significant when the magnetic field is weak, such that ICS dominates the electron energy loss.

\subsection{\texttt{Aniso\_A} model}
\label{app:aniso}
Anisotropic diffusion can be solved in cylindrical coordinates through a coordinate transformation \cite{Fang:2023axu}. Given that $D_{rr}=M_A^4D_{zz}$, if we define a new coordinate $z'$ such that $z'=M_A^2z$, the diffusion equation becomes isotropic in the $r-z'$ coordinate system. Based on Eqs.~(\ref{eq:green}) and (\ref{eq:green_norm}), it is straightforward to derive its solution in cylindrical coordinates. Subsequently, by performing a variable substitution, we can obtain a semi-analytical solution for anisotropic diffusion, as
\begin{equation}
 N(E_e, r, z, t_s)=\int_{t_\mathrm{min}}^{t_s}dt_0\,q_t(t_0)\,q_E(E_{e,\star})\,\frac{b(E_{e,\star})}{b(E_e)}\,\frac{M_A^2}{(4\pi\lambda_\star)^{3/2}}\,\exp\left(-\frac{r^2+M_A^4z^2}{4\lambda_\star}\right)\,,
 \label{eq:sol_ani}
\end{equation}
where
\begin{equation}
 \lambda_\star=\int_{E_e}^{E_{e,\star}}\frac{D_{rr}(E_e^\prime)}{b(E_e^\prime)}dE_e^\prime\,.
 \label{eq:lambda_ani}
\end{equation}

\subsection{\texttt{Super\_A} model}
\label{app:super}
For the fractional diffusion equation, we can extend the Green's function as follows (see, e.g., Ref.~\cite{2001NuPhS..97..267L}):
\begin{equation}     
G(E_e,\mathbi{r},t;E_{e,0},\mathbi{r}_0,t_0)=\frac{\rho_3^{(\alpha)}(|\mathbi{r}-\mathbi{r}_0|\lambda^{-1/\alpha})}{b(E_e)\lambda^{3/\alpha}}\,\delta(t-t_0-\tau)H(\tau)\,,
 \label{eq:green_sup}
\end{equation}
where
\begin{equation}
 \rho_3^{(\alpha)}(r)=\frac{1}{2\pi^2r}\int_0^\infty\exp\left(-k^\alpha\right)\,\sin(kr)kdk\,.
 \label{eq:rho3}
\end{equation}
The solution for the \texttt{Super\_A} model is then
\begin{equation}
 N(E_e, r, t_s)=\int_{t_\mathrm{min}}^{t_s}dt_0\,q_t(t_0)\,q_E(E_{e,\star})\,\frac{b(E_{e,\star})}{b(E_e)}\,\frac{\rho_3^{(\alpha)}(r\lambda_\star^{-1/\alpha})}{\lambda_\star^{3/\alpha}}\,.
 \label{eq:sol_sup}
\end{equation}
When $\alpha=2$, $\rho_3^{(2)}(r)=1/(4\pi)^{3/2}\exp(-r^2/4)$, the solution of superdiffusion reduces to Eq.~(\ref{eq:sol_norm}).

\subsection{\texttt{Juttner\_A} model}
\label{app:normj}
J\"{u}tnner introduced a relativistic correction to the Maxwell–Boltzmann distribution. Due to the formal similarity between the solution kernel of the diffusion equation and the Maxwell–Boltzmann distribution, Ref.~\cite{Aloisio:2008tx} employed an analogous method to address the superluminal problem in the diffusion equation. For the diffusion-loss equation of electrons, the corrected Green's function is expressed as
\begin{equation}
 \begin{aligned}
  G(E_e,\mathbi{r},t;E_{e,0},\mathbi{r}_0,t_0) =& \frac{1}{b(E_e)4\pi[c(t-t_0)]^3}\,\frac{H(1-\xi)}{(1-\xi^2)^2}\,\frac{\kappa}{K_1(\kappa)}\exp\left(-\frac{\kappa}{\sqrt{1-\xi^2}}\right) \\
  & \times\delta(t-t_0-\tau)H(\tau)\,, \\
 \end{aligned}
 \label{eq:green_juttner}
\end{equation}
where $K_1$ is the first-order modified Bessel function of the second kind, $\xi=(\mathbi{r}-\mathbi{r}_0)/c(t-t_0)$, and $\kappa=[c(t-t_0)]^2/2\lambda$. The solution is thus
\begin{equation}
 \begin{aligned}
  N(E_e, r, t_s)=&\int_{t_\mathrm{min}}^{t_s}dt_0\,q_t(t_0)\,q_E(E_{e,\star})\,\frac{b(E_{e,\star})}{b(E_e)}\,\frac{1}{4\pi[c(t_s-t_0)]^3}\,\frac{H(1-\xi)}{(1-\xi^2)^2}\,\frac{\kappa}{K_1(\kappa)} \\
  & \times\exp\left(-\frac{\kappa}{\sqrt{1-\xi^2}}\right)\,, \\
 \end{aligned}
 \label{eq:sol_juttner}
\end{equation}
where $\xi=r/c(t_s-t_0)$, and $\kappa=[c(t_s-t_0)]^2/2\lambda_\star$.

Furthermore, when the electron propagation is quasi-ballistic, the velocity distribution of electrons exhibits anisotropy, and the formula for describing isotropic ICS (Eq.~(\ref{eq:sg})) is no longer valid. However, when performing the LOS integration, we can equivalently transfer the ICS correction to the electron number density:
\begin{equation}
 S_e(E_e,\theta)=\int_0^\infty 2N(E_e, r)M(x,\mu)dl\,,
 \label{eq:s_juttner}
\end{equation}
where the correction factor, as suggested by Ref.~\cite{Prosekin:2015ima}, takes the form of
\begin{equation}
 M(x, \mu)=\frac{1}{Z(x)}\,\exp\left[-\frac{3(1-\mu)}{x}\right]\,,
 \label{eq:M}
\end{equation}
where $x=rc/D(E_e)$, $Z(x)=\frac{x}{3}\left[1-\exp\left(-\frac{6}{x}\right)\right]$, and $\mu=(\mathtt{rs}\cos\theta-l)/r$. After this revision, we can still use Eq.~(\ref{eq:sg}) for the ICS computation.

It should be noted that for slow diffusion, quasi-ballistic propagation occurs only at extremely small spatial and temporal scales. Consequently, the corrections introduced by Eqs.~(\ref{eq:sol_juttner}) and (\ref{eq:s_juttner}) are negligible, and the solution reduces to that of the \texttt{StdDiff\_A} model.

\section{Numerical solutions to electron propagation}
\label{app:numr}
Equation~(\ref{eq:prop}) can be rewritten as
\begin{equation}
 \frac{\partial N}{\partial t}=\mathcal{L}N+Q\,.
 \label{eq:prop_num}
\end{equation}
For spherically symmetric models, $\mathcal{L} = \mathcal{L}_E + \mathcal{L}_r$, where $\mathcal{L}_E$ is the energy-loss operator, and $\mathcal{L}_r$ is the diffusion operator in the $r$ direction in spherical coordinates. For cylindrically symmetric models, $\mathcal{L} = \mathcal{L}_E + \mathcal{L}_r + \mathcal{L}_z$, where $\mathcal{L}_r$ ($\mathcal{L}_z$) is the the diffusion operator in the $r$ ($z$) direction in cylindrical coordinates.

We employ the operator splitting method to solve the problem, which allows us to manage each operator in the propagation equation independently. Specifically, we employ Strang splitting with second-order time accuracy \cite{1968SJNA....5..506S}. For the spherically symmetric scenario, the procedure of Strang splitting is as follows:
\begin{equation}
N^{n+1}=\mathcal{S}_E(t^{n+1/2},t^{n+1})\mathcal{S}_r(t^{n},t^{n+1})\mathcal{S}_E(t^{n},t^{n+1/2})N^{n}\,,
\label{eq:strang1}
\end{equation}
where $t^{n+1/2}=t^n+\Delta t/2$, and $\mathcal{S}_*$ is the numerical method for updating $N$ within the subproblem $\mathcal{L}_*$. The source function is integrated into $\mathcal{S}_E$. Similarly, the Strang splitting is as follows for the cylindrically symmetric scenario:
\begin{equation}
N^{n+1}=\mathcal{S}_E(t^{n+1/2},t^{n+1})\mathcal{S}_z(t^{n+1/2},t^{n+1})\mathcal{S}_r(t^{n},t^{n+1})\mathcal{S}_z(t^{n},t^{n+1/2})\mathcal{S}_E(t^{n},t^{n+1/2})N^{n}\,.
\label{eq:strang2}
\end{equation}
Note that in Eq.~(\ref{eq:strang1}) and (\ref{eq:strang2}), the operator closer to $N^n$ acts on $N^n$ first.

As the energy-loss and diffusion operators do not commute, if a sequential splitting is adopted for operator splitting, the time discretization error is first order. The Strang splitting symmetrically divides each time step, allowing the non-commuting operators to act in an alternating sequence. This cancels the $O(\Delta t)$ term in the truncation error, raising the time accuracy to second order. According to our tests, to achieve the same level of accuracy, sequential splitting requires a time step more than an order of magnitude smaller than that needed for Strang splitting. Therefore, Strang splitting ensures significantly higher computational efficiency.


In general, we use a constant time step, i.e., $\Delta t=\mathtt{dt0}$. As detailed below, we adopt the Crank–Nicolson scheme for time integration. Although this scheme provides second‑order temporal accuracy and is unconditionally stable, it is not L‑stable \cite{Hairer96}. The point-source diffusion we address is a stiff problem; for schemes that lack L‑stability, if the time step $\Delta t$ is not sufficiently small, the solution may exhibit strong oscillations near the source. Therefore, as $t$ approaches $t_s$, we take several small time steps of $\Delta t=1$~yr to eliminate the oscillations. In spherically symmetric geometries, two such steps are enough to effectively damp the oscillations, whereas in cylindrically symmetric cases we employ 10 steps. This strategy allows larger time steps to be used over most of the evolution, maintaining computational efficiency.


To derive the discretization for the energy-loss process (including injection), we adopt the integration method given in Ref.~\cite{Kissmann:2014sia}. We integrate $\partial N/\partial t = \partial (bN)/\partial E_e + Q$ over the rectangular region $[E_{e,j}, E_{e,j-1}] \times [t^n, t^{n+1}]$ using the trapezoidal rule, resulting in:
\begin{equation}
 \begin{aligned}
  & (\Delta E_e-b_{j-1}\Delta t)N_{j-1}^{n+1}+(\Delta E_e+b_j\Delta t)N_j^{n+1} \\
  = & (\Delta E_e+b_{j-1}\Delta t)N_{j-1}^n+(\Delta E_e-b_j\Delta t)N_j^n \\ 
  & + \frac{\Delta E_e\Delta t}{2}\left(Q^{n+1}_{j-1}+Q^{n+1}_{j}+Q^{n}_{j-1}+Q^{n}_{j}\right) \\
 \end{aligned}
 \,,
 \label{eq:scheme_box}
\end{equation}
where $\Delta E_e=E_{e,j-1}-E_{e,j}$, and the superscripts (subscripts) represent the time (energy) dimension. Note that $E_{e,j}$ is arranged in descending order, i.e., $\Delta E_e>0$. The advantages of this scheme include: 1) inherent flux conservation; 2) unconditional stability; 3) second-order accuracy. Since the energy-loss process involves energy flowing from high to low, we assume a high-energy boundary condition $N(\mathtt{Ee\_max},t)=0$. Therefore, despite the fact that the form of Eq.~(\ref{eq:scheme_box}) is implicit, the terms $N_{j-1}^{n+1}$, $N_{j-1}^n$, and $N_j^n$ are all known, allowing us to explicitly derive $N_j^{n+1}$.

It should be clarified that \texttt{Ee\_max}, by definition, is the maximum electron energy \textit{used in computations}, rather than the maximum energy of the electron injection spectrum itself. However, given the boundary condition $N(\mathtt{Ee\_max},t)=0$ in the numerical models, \texttt{Ee\_max} essentially serves as the upper limit of the integral in Eq.~(\ref{eq:eta}). Therefore, to ensure the effectiveness of the cutoff energy \texttt{Ec} for the electron spectrum, it is recommended to set \texttt{Ee\_max} significantly larger than \texttt{Ec}.

For the diffusion operators, we use the finite volume method (FVM) for discretization \cite{LeVeque-2002}. In the following, we separately present the scenarios of spherically symmetric diffusion and cylindrically symmetric diffusion.

\subsection{Spherically symmetric diffusion}
\label{app:sph}
We divide the space into spherical shells with inner and outer radii of $r_{j-1/2}$ and $r_{j+1/2}$, respectively, where $j=0,1,2,...,J$\footnote{Since the number of spatial grid points is \texttt{NUM\_r}, we have $J=\mathtt{NUM\_r-1}$.}, referred to as control volumes (CVs). The principle of FVM is to integrate the equation over each CV and transform the volume integral into the flux difference across the interfaces, ensuring that the temporal variation of conserved quantities within each CV balances the net flux through its boundaries. For non-uniform spatial grids (e.g., Eq.~(\ref{eq:grid_r})) or discontinuous coefficients (e.g., Eq.~(\ref{eq:d_2z})), FVM can directly construct discrete schemes with good conservation properties, which is generally difficult to achieve with conventional finite difference methods.

Integrating the equation $\partial N/\partial t=\mathcal{L}_rN$ over the spherical shell $[r_{j-1/2},r_{j+1/2}]$, we obtain
\begin{equation}
 \begin{aligned}
  V_j\frac{d\bar{N}_j}{dt} & =\int_{V_j}\nabla\cdot(D\nabla N)dV=\oint_{\partial V_j}D\nabla N\cdot \mathbi{n}dA \\
  & =4\pi(r_{j+1/2}^2F_{j+1/2}-r_{j-1/2}^2F_{j-1/2})\,,
 \end{aligned}
 \label{eq:fvm_intg}
\end{equation}
where $V_j=\frac{4\pi}{3}(r_{j+1/2}^3-r_{j-1/2}^3)$ is the CV size, $\mathbi{n}$ is the normal vector of the CV interface, and $F\equiv D\nabla N$ is the flux density through the interface.

It should be noted that $\bar{N}_j$ represents the average value of $N$ within the control volume, i.e., 
\begin{equation}
 \bar{N}_j=\frac{1}{V_j}\int_{V_j}\frac{\partial N}{\partial t}dV\,.
 \label{eq:fvm_nbar}
\end{equation}
Therefore, Eq.~(\ref{eq:fvm_intg}) is exact. FVM approximates the true value of $N$ at the center of the CV $r_j=(r_{j+1/2}+r_{j-1/2})/2$ using $\bar{N}_j$, which fundamentally differs from finite difference methods.

Below, we discretize $F_{j+1/2}$ and $F_{j-1/2}$. Since $D$ and $\nabla N$ may exhibit discontinuities at the interface, it is necessary to discretize the flux on each side of the interface separately. Taking $F_{j+1/2}$ as an example, we have
\begin{equation}
 \begin{aligned}
 & F_{j+1/2}^-=D\left.\frac{\partial N}{\partial r}\right|_{r_{j+1/2}^-}\approx D_j\frac{N_{j+1/2}^--\bar{N}_j}{\Delta r_j/2} \\
 & F_{j+1/2}^+=D\left.\frac{\partial N}{\partial r}\right|_{r_{j+1/2}^+}\approx D_{j+1}\frac{\bar{N}_{j+1}-N_{j+1/2}^+}{\Delta r_{j+1}/2} \\
 \end{aligned}
 \,,
 \label{eq:fvm_single}
\end{equation}
where $\Delta r_j=r_{j+1/2}-r_{j-1/2}$ is the CV thickness. By using the continuity of particle number ($N_{j+1/2}^-=N_{j+1/2}^+=N_{j+1/2}$) and flux density ($F_{j+1/2}^-=F_{j+1/2}^+=F_{j+1/2}$), we obtain
\begin{equation}
 F_{j+1/2}=\frac{\bar{N}_{j+1}-\bar{N}_j}{\frac{1}{2}\left(\frac{\Delta r_{j+1}}{D_{j+1}}+\frac{\Delta r_j}{D_j}\right)}\equiv\frac{\bar{N}_{j+1}-\bar{N}_j}{R_{j+1/2}}\,,
 \label{eq:fvm_flux}
\end{equation}
where $R_{j+1/2}$ can be defined as the diffusion resistance. The greater the diffusion resistance, the smaller the interface flux. It is evident that $R$ tends to be determined by the side with the smaller diffusion coefficient and the greater CV thickness. Similarly, 
\begin{equation}
 F_{j-1/2}=\frac{\bar{N}_{j}-\bar{N}_{j-1}}{\frac{1}{2}\left(\frac{\Delta r_j}{D_j}+\frac{\Delta r_{j-1}}{D_{j-1}}\right)}\equiv\frac{\bar{N}_j-\bar{N}_{j-1}}{R_{j-1/2}}\,.
 \label{eq:fvm_flux2}
\end{equation}

Combining Eqs.~(\ref{eq:fvm_intg}), (\ref{eq:fvm_flux}), and (\ref{eq:fvm_flux2}), we obtain the semi-discrete scheme of the diffusion equation:
\begin{equation}
 \frac{d\bar{N}_j}{dt}=\frac{4\pi}{V_j}\left(r_{j+1/2}^2\frac{\bar{N}_{j+1}-\bar{N}_j}{R_{j+1/2}}-r_{j-1/2}^2\frac{\bar{N}_j-\bar{N}_{j-1}}{R_{j-1/2}}\right)\,.
 \label{eq:fvm_semi}
\end{equation}
Performing trapezoidal integration on the Eq.~(\ref{eq:fvm_semi}) over $[t_n, t_{n+1}]$, a Crank-Nicolson type implicit scheme is obtained. After a simple rearrangement, the final discretization is
\begin{equation}
 -L_j\bar{N}_{j-1}^{n+1}+(1+L_j+U_j)\bar{N}_j^{n+1}-U_j\bar{N}_{j+1}^{n+1}=
 L_j\bar{N}_{j-1}^n+(1-L_j-U_j)\bar{N}_j^n+U_j\bar{N}_{j+1}^n\,,
 \label{eq:fvm_sph}
\end{equation}
where
\begin{equation}
 L_j=\frac{2\pi r_{j-1/2}^2\Delta t}{V_jR_{j-1/2}}\,,\quad U_j=\frac{2\pi r_{j+1/2}^2\Delta t}{V_jR_{j+1/2}}\,.
 \label{eq:fvm_sph_LU}
\end{equation}

Specifically, when $j=0$, CV is no longer a spherical shell but a sphere with a radius of $\Delta r_0 = r_{1/2}$. At this point, CV has only an outer interface and no inner interface, and the semi-discrete scheme is written as
\begin{equation}
 \frac{d\bar{N}_0}{dt}=\frac{4\pi r_{1/2}^2}{V_0}\frac{\bar{N}_1-\bar{N}_0}{R_{1/2}}\,,
 \label{eq:fvm_semi0}
\end{equation}
where $V_0=\frac{4\pi}{3}(\Delta r_0)^3$, and $R_{1/2}=\frac{\Delta r_1}{2D_1}+\frac{\Delta r_0}{D_0}$. The corresponding final discretization is
\begin{equation}
 (1+U_0)\bar{N}_0^{n+1}-U_0\bar{N}_1^{n+1}=
 (1-U_0)\bar{N}_0^n+U_0\bar{N}_1^n\,.
 \label{eq:fvm_sph0}
\end{equation}
Moreover, assuming the outer boundary condition is $F_{J+1/2}=0$, the discretization at the outer boundary is given by
\begin{equation}
 -L_J\bar{N}_{J-1}^{n+1}+(1+L_J)\bar{N}_J^{n+1}=
 L_J\bar{N}_{J-1}^n+(1-L_J)\bar{N}_j^n\,.
 \label{eq:fvm_sphJ}
\end{equation}

Equations~(\ref{eq:fvm_sph}), (\ref{eq:fvm_sph0}), and (\ref{eq:fvm_sphJ}) form a tridiagonal system:
\begin{equation}
 \begin{pmatrix}
  1+U_0 & -U_0 & & & \\
  -L_1 & 1+L_1+U_1 & -U_1 & & \\
  & \ddots & \ddots & \ddots & \\
  & & -L_{J-1} & 1+L_{J-1}+U_{J-1} & -U_{J-1} \\
  & & & -L_J & 1+L_J \\
 \end{pmatrix}
 \begin{pmatrix}
  \bar{N}_0^{n+1} \\
  \bar{N}_1^{n+1} \\
  \vdots \\
  \bar{N}_{J-1}^{n+1} \\
  \bar{N}_J^{n+1} \\
 \end{pmatrix}
 =
 \begin{pmatrix}
  f_0 \\
  f_1 \\
  \vdots \\
  f_{J-1} \\
  f_J \\
 \end{pmatrix}
 \,,
 \label{eq:fvm_tridag}
\end{equation}
where
\begin{equation}
 \begin{aligned}
  & f_0=(1-U_0)\bar{N}_0^n+U_0\bar{N}_1^n\,,\\
  & f_j=L_j\bar{N}_{j-1}^n+(1-L_j-U_j)\bar{N}_j^n+U_j\bar{N}_{j+1}^n \,,\quad j=1,2,...,J-1\,,\\
  & f_J=L_J\bar{N}_{J-1}^n+(1-L_J)\bar{N}_J^n\,.\\
 \end{aligned}
 \label{eq:fvm_f}
\end{equation}
Equation (\ref{eq:fvm_tridag}) can be solved by the Thomas method, thereby updating $\bar{N}^n$ to $\bar{N}^{n+1}$.

Note that under FVM, the boundaries of each CV are determined first, followed by defining the CV center. Consequently, the spatial grid setting used in the numerical solutions of \texttt{PHECT} differs from that described in Eq.~(\ref{eq:grid_r}). When $j\neq0$, 
\begin{equation}
 \begin{aligned}
  & r_{j-1/2}=\frac{dr}{a}\tan[a(j-1)]+\frac{dr}{2}\,, \\
  \Rightarrow\,\,& r[j]=\frac{r_{j-1/2}+r_{j+1/2}}{2}=\frac{dr}{a}\left\{\frac{\tan[a(j-1)]+\tan(aj)}{2}\right\}+\frac{dr}{2}\,.\\
 \end{aligned}
 \label{eq:fvm_grid_r}
\end{equation}
Specifically, $r_0=0$. From Eq.~(\ref{eq:fvm_grid_r}), it can be seen that when $j$ is small, $r[j]\approx jdr$, which is close to Eq.~(\ref{eq:grid_r}). However, as $j$ increases, $r[j]$ gradually becomes smaller than $dr/a\tan(aj)$. In fact, \texttt{r\_max} defined in \texttt{param\_config.yaml} approximately corresponds to the outer boundary of the outermost CV, being larger than the center of that CV.

For the source function, we express it as $Q(E,r,t)=q_E(E)q_r(r)q_t(t)$ as introduced in Sec.~\ref{subsubsec:common}. In Eq.~(\ref{eq:src_r}), a point source assumption is used, while in the numerical computation of the spherically symmetric models, it is assumed that the source function is uniformly distributed within the CV, $V_0$. Thus, the spatial component of the source function is given as
\begin{equation}
 q_r(r)=\left\{
 \begin{aligned}
  & \frac{1}{\frac{4\pi}{3}(dr/2)^3}\,, \quad & r\in V_0 \\
  & 0\,, \quad & r\notin V_0 \\
 \end{aligned}
 \right.\,.
 \label{eq:fvm_src_r}
\end{equation}

\subsection{Cylindrically symmetric diffusion}
\label{app:cylin}
Within the framework of the operator splitting method, we discretize the 2D radially symmetric diffusion in the $r$ direction and the 1D diffusion in the $z$ direction separately. The derivation process is similar to that in Appendix~\ref{app:sph}, and we present the results directly below.

The semi-discrete scheme for radially symmetric diffusion is 
\begin{equation}
 \frac{d\bar{N}_j}{dt}=\frac{2\pi}{A_j}\left(r_{j+1/2}\frac{\bar{N}_{j+1}-\bar{N}_j}{R_{j+1/2}}-r_{j-1/2}\frac{\bar{N}_j-\bar{N}_{j-1}}{R_{j-1/2}}\right)\,,
 \label{eq:fvm_cylin_semi1}
\end{equation}
where $A_j=\pi(r_{j+1/2}^2-r_{j-1/2}^2)$. The final discretization is the same as Eq.~(\ref{eq:fvm_sph}), but
\begin{equation}
 L_j=\frac{\pi r_{j-1/2}\Delta t}{A_jR_{j-1/2}}\,,\quad U_j=\frac{\pi r_{j+1/2}\Delta t}{A_jR_{j+1/2}}\,.
 \label{eq:fvm_cylin_LU1}
\end{equation}
Additionally, the special treatment of boundary conditions is the same as that discussed in Appendix~\ref{app:sph} and is thus not elaborated on further.

The semi-discrete scheme for the diffusion in the $z$ direction is
\begin{equation}
 \frac{d\bar{N}_j}{dt}=\frac{1}{\Delta z_j}\left(\frac{\bar{N}_{j+1}-\bar{N}_j}{R_{j+1/2}}-\frac{\bar{N}_j-\bar{N}_{j-1}}{R_{j-1/2}}\right)\,,
 \label{eq:fvm_cylin_semi2}
\end{equation}
where
\begin{equation}
 R_{j+1/2}=\frac{1}{2}\left(\frac{\Delta z_{j+1}}{D_{j+1}}+\frac{\Delta z_j}{D_j}\right)\,,\quad 
 R_{j-1/2}=\frac{1}{2}\left(\frac{\Delta z_j}{D_j}+\frac{\Delta z_{j-1}}{D_{j-1}}\right)\,.
 \label{eq:fvm_z_R}
\end{equation}
The final discretization is the same as Eq.~(\ref{eq:fvm_sph}), but
\begin{equation}
 L_j=\frac{\Delta t}{2\Delta z_jR_{j-1/2}}\,,\quad 
 U_j=\frac{\Delta t}{2\Delta z_jR_{j+1/2}}\,.
 \label{eq:fvm_cylin_LU2}
\end{equation}
The special treatment of boundary conditions is not elaborated on further.

For the \texttt{SNR2Z\_N} model, we refer Eq.~(\ref{eq:d_snr}) to set the spacially dependent $D$. For the \texttt{Aniso\_N} model, we set constant diffusion coefficient $D_{rr}$ ($D_{zz}$) in the $r$ ($z$) direction scheme.

The grid setting in the $z$ direction is symmetric about $z=0$. From $z=0$ to $z=\mathtt{z\_max}$, there are a total of \texttt{NUM\_z} grid points, thus the actual array size in the $z$ direction is $2\mathtt{NUM\_z}-1$. Specifically, the values of the grid points are given by:
\begin{equation}
 z[j]=\left\{
 \begin{aligned}
 & \frac{dz}{a}\left\{\frac{\tan[a(j-\mathtt{NUM\_z})]+\tan\left[a(j-\mathtt{NUM\_z}+1\right]}{2}\right\}+\frac{dz}{2}\,,& \quad j>\mathtt{NUM\_z-1} \\
 & 0 \,,\quad & j=\mathtt{NUM\_z-1} \\
 & z[2\mathtt{NUM\_z}-2-j] \,,& \quad j<\mathtt{NUM\_z-1} \\
 \end{aligned}
 \right.\,,
 \label{eq:fvm_grid_z}
\end{equation}
where the determinations for $a$ and $dz$ are analogous to those Eq.~(\ref{eq:grid_r2}).

Assuming that the source function is uniformly distributed within the innermost cylindrical CV, $V_0$, around the coordinate origin, the spatial term is then expressed as
\begin{equation}
 q_r(r,z)=\left\{
 \begin{aligned}
  & \frac{1}{\pi(dr/2)^2dz}\,, \quad & (r,z)\in V_0 \\
  & 0\,, \quad & (r,z)\notin V_0 \\
 \end{aligned}
 \right.\,.
 \label{eq:fvm_src_rz}
\end{equation}

\section{Computation time}
\label{app:time}
Table~\ref{tab:time} presents the time needed for a full computation for each model. The tests are conducted on a workstation equipped with an Intel Core i9-14900HX CPU and 62 GB of RAM. Default parameters presented in Table~\ref{tab:params} are used in the tests.

The runtime of \texttt{Aniso\_A} appears to be remarkably slower than other models. As introduced in Appendix~\ref{app:ana}, its computational logic is identical to that of \texttt{StdDiff\_A}, while it includes an additional spatial dimension as a cylindrically symmetric model. With the default setting $\texttt{NUM\_z}=60$, the runtime of \texttt{Aniso\_A} is roughly several tens of times longer than that of \texttt{StdDiff\_A}.

\begin{table}[!htb]
 \centering
 \caption{Runtime of different models} \label{tab:time}
 \begin{tabular}{lc}
  \toprule
  Model & Real time \\ 
  \midrule
  \texttt{StdDiff\_N} & $\approx2.5$s \\
  \texttt{StdDiff\_A} & $\approx50$s \\
  \texttt{StdDiff\_A0} & $\approx1.5$s \\
  \texttt{Juttner\_A} & $\approx50$s \\
  \texttt{Super\_A} & $\approx50$s \\
  \texttt{2Zone\_N} & $\approx2.5$s \\
  \texttt{SNR2Z\_N} & $\approx3$m$40$s \\
  \texttt{Aniso\_N} & $\approx3$m$30$s \\
  \texttt{Aniso\_A} & $\approx47$m \\
  \bottomrule
 \end{tabular}
\end{table}

\section{Corrections to EDGE}
\label{app:edge}
In Sec.~\ref{sec:test}, we mentioned that a significant issue with \texttt{EDGE} is that it does not reset the background photon field for ICS during each initialization. Below is a snippet of the original initialization program:
\begin{lstlisting}[language=Python]
def InitialiseGappa(fp,fr,b,age):
    fr.AddThermalTargetPhotons(2.7,0.26*gp.eV_to_erg)   #CMB
    fr.AddThermalTargetPhotons(TIR,WIR*gp.eV_to_erg)   #IR
    fr.AddThermalTargetPhotons(TOPT,WOPT*gp.eV_to_erg) #OPT
\end{lstlisting}
where \texttt{AddThermalTargetPhotons} accumulates the background photon field.

\begin{figure}[!htb]
\centering
\includegraphics[width=0.5\textwidth]{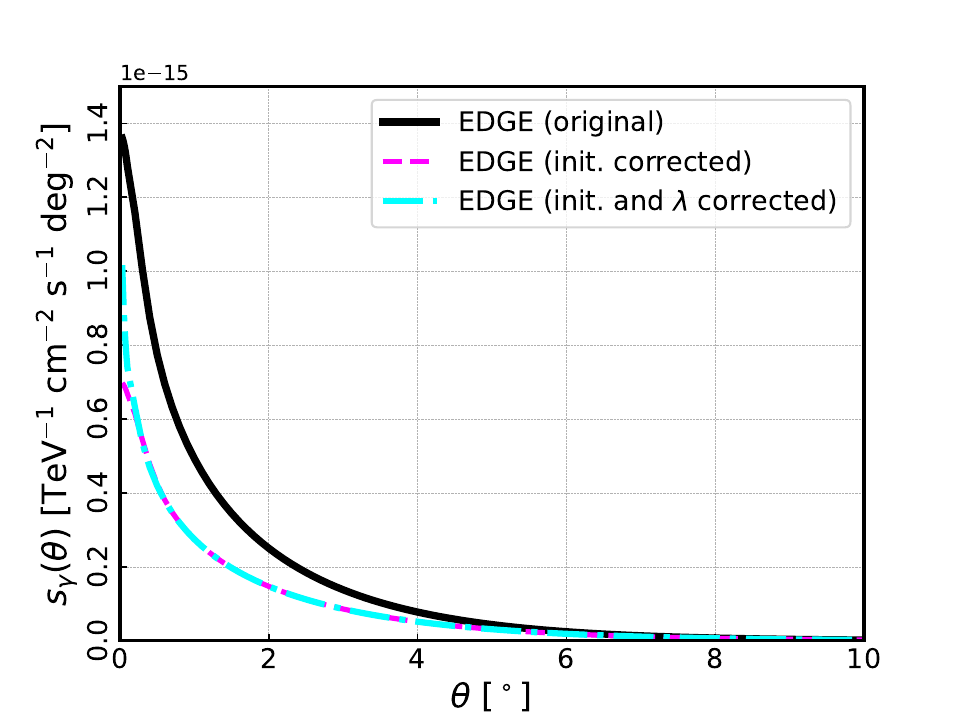}
\caption{Energy differential 1D $\gamma$SB of the Geminga halo computed by \texttt{EDGE}, including the original result, the result after correcting the background photon initialization issue (init. corrected), and the further improved results after correcting the calculation of $\lambda$ (init. and $\lambda$ corrected).}
\label{fig:edge_app}
\end{figure}

Subsequently, the background photon field is initialized twice in the main function, as shown in the code snippet below:
\begin{lstlisting}[language=Python]
    fp = InitialiseGappa(fp,fr,BBURST,AGEBURST)
    ETRAJBURST,ETRAJBURSTINVERSE,LAMBBURST = CalculateEnergyTrajectory(fp)
    fp = InitialiseGappa(fp,fr,BCONT,AGECONT)
    ETRAJCONT,ETRAJCONTINVERSE,LAMBCONT = CalculateEnergyTrajectory(fp)
\end{lstlisting}
Although these initializations are intended for different calculation branches, the background photon field has already been summed twice at this point, resulting in an overestimation of the final $\gamma$-ray flux. As shown in Fig.~\ref{fig:edge_app}, after correcting this issue (e.g., by deactivating the ``BURST'' branch), the $\gamma$-ray flux reduces by approximately half.

The magenta dashed line in Fig.~\ref{fig:edge_app} shows the 1D $\gamma$SB after correcting the initialization problem of photon fields. Nonetheless, a notable issue persists: the gradient gradually decreases as $\theta$ approaches $0^\circ$. For continuous electron injection, the expected $\gamma$SB distribution at small angles should be sharp because the newly injected electrons do not diffuse sufficiently and tend to accumulate near the source. Since \texttt{EDGE} does not perform PSF convolution, the profile should not flatten at small angles. The cause lies in the fact that \texttt{EDGE} overestimates the characteristic diffusion distance $\sqrt{\lambda_\star}$ of the recently injected electrons.

\texttt{EDGE} estimates $\lambda_\star$ using a table lookup method. For each given electron energy $E_e$, \texttt{EDGE} prepares a set of $E_{e,0}$ values along with the corresponding $\lambda$ values for table lookup (refer to the meaning in Eq.~(\ref{eq:lambda})). During the time integration in Eq.~(\ref{eq:sol_norm}), \texttt{EDGE} first determines the $E_{e,\star}$ corresponding to $t_0$, and then interpolates to obtain $\lambda_\star$ from the prepared $(E_{e,0},\lambda)$ pairs.

\begin{figure}[!htb]
\centering
\includegraphics[width=0.45\textwidth]{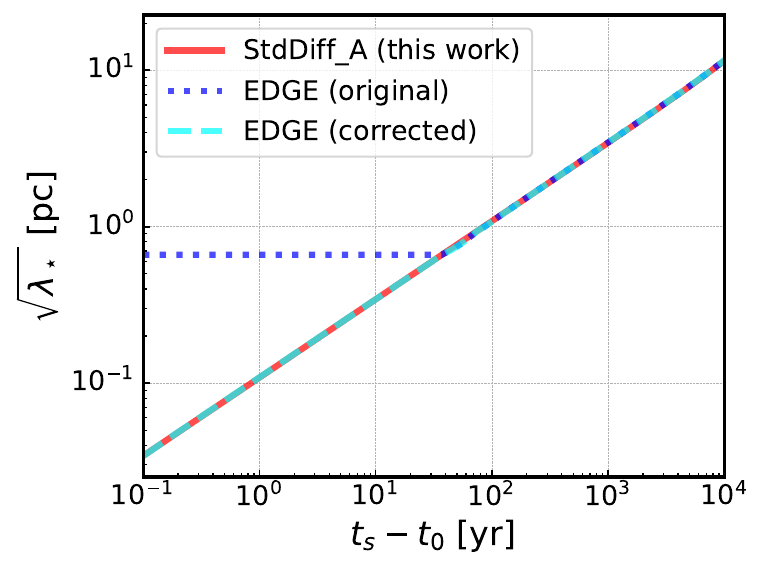}
\includegraphics[width=0.45\textwidth]{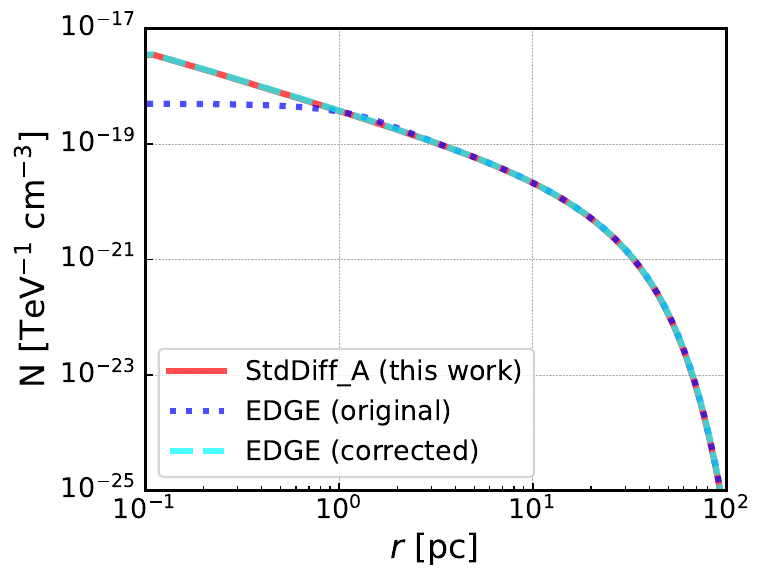}
\caption{Changes introduced by correcting the electron characteristic diffusion distance $\sqrt{\lambda_\star}$ in the \texttt{EDGE} code, along with a comparison to results from the \texttt{StdDiff\_A} model in \texttt{PHECT}. The left panel shows $\sqrt{\lambda_\star}$ as a function of electron injection time, and the right panel presents the spatial distribution of electron number density. The electron energy is fixed at $50$~TeV.}
\label{fig:lambda}
\end{figure}

For each $E_e$, \texttt{EDGE} logarithmically divides the range of $(E_e,E_\mathrm{MAX})$ to obtain the $E_{e,0}$ grid points. However, $E_{e,\star}$ does not have a linear relationship with $t_0$. When $E_{e,\star}$ approaches $E_e$, the variation of $t_0$ with respect to $E_{e,\star}$ is highly sensitive. In other words, as $t_0$ approaches $t_s$, the difference between $E_{e,\star}$ and $E_e$ becomes extremely small, potentially smaller than the resolution of the $E_{e,0}$ grid. In such cases, \texttt{EDGE} directly uses the $\lambda$ corresponding to the smallest $E_{e,0}$ grid point to estimate $\lambda_\star$, leading to an overestimation of the electron diffusion distance, as shown in the left panel of Fig.~\ref{fig:lambda}. This results in a flattening of the electron number density at small $r$ (the right panel of Fig.~\ref{fig:lambda}), consequently affecting the morphology of $\gamma$SB at small angles.

To address this issue, we densely sample $E_{e,0}$ in the vicinity of $E_{e}$ for the \texttt{EDGE} code. This adjustment results in a revised $\lambda_\star$ that is consistent with the results of \texttt{PHECT}, thereby achieving a consistency in the electron density distribution as illustrated in Fig.~\ref{fig:lambda}.

Additionally, \texttt{EDGE} has several other issues that could affect the accuracy of the calculations. When calculating the electron number density, the factor $b_0(E_{e,\star})/b_0(E_e)$ in Eq.~(\ref{eq:sol_norm}) is neglected\footnote{Correcting this issue for \texttt{EDGE} may be tedious, so we applied the same disregard to \texttt{PHECT} when comparing with \texttt{EDGE}.}. When performing the time integration of electron number density, according to its original convention, the integration range should be $[\max(\texttt{tmin}, \texttt{AGECONT-DT}),\texttt{AGECONT-1e-3}]$ instead of $[\max(\texttt{1e-3}, \texttt{AGECONT-DT}),\texttt{AGECONT-tmin}]$. In the line-of-sight integration, the original code directly uses the number density at the nearest $r$ grid point to determine the electron number density along the path, without employing interpolation.

\bibliography{references}

\end{document}